\documentclass[aps,twocolumn,prd,superscriptaddress,nofootinbib]{revtex4-1}

\usepackage[utf8]{inputenc}

\usepackage{mathtools}
\usepackage{amsfonts}
\usepackage{mathrsfs}
\usepackage{bbm}
\usepackage{slashed}

\usepackage{graphicx}
\usepackage{subcaption}
\usepackage{color}
\usepackage{array}

\usepackage{placeins}
\usepackage{booktabs}
\usepackage{caption}

\usepackage{xspace}
\usepackage{siunitx}
\usepackage{hyperref}
\usepackage[nameinlink]{cleveref}
\usepackage{bookmark}

\usepackage{xifthen}
\usepackage{xcolor}
\hypersetup{
	colorlinks,
	linkcolor={red!75!black},
	citecolor={blue!75!black},
	urlcolor={blue!75!black}
}
\usepackage{units}

\setkeys{Gin}{width=0.48\textwidth}
\newcommand{\imag}{\text{i}}

\newcommand{\tinytext}[1]{\text{\tiny{#1}}}

\graphicspath{{./figures/}}

\newcommand{\gettitle}{Bound state properties from the Functional Renormalisation Group}

\hypersetup{
	colorlinks,
	linkcolor={red!75!black},
	citecolor={blue!75!black},
	urlcolor={blue!75!black},
	pdftitle={\gettitle},
	pdfauthor={Alkofer, Maas, Mian, Mitter, Paris-Lopez, Pawlowski, Wink},
	pdfkeywords={effective theory} {analytic continuation}
	{correlations functions} {hadronization}
	{functional renormalisation group}
	{real time} {bound states}
	bookmarksopen=true,
	bookmarksopenlevel=2,
	bookmarksnumbered=true
}

\begin{document}

\title{\gettitle}

\author{Reinhard Alkofer}
  \affiliation{Institute of Physics, NAWI Graz, University of Graz, Universit\"atsplatz 5, 8010 Graz, Austria} 

\author{Axel Maas}
  \affiliation{Institute of Physics, NAWI Graz, University of Graz, Universit\"atsplatz 5, 8010 Graz, Austria} 

\author{Walid Ahmed Mian}
  \affiliation{Institute of Physics, NAWI Graz, University of Graz, Universit\"atsplatz 5, 8010 Graz, Austria} 
  \affiliation{Institut f\"ur Theoretische Physik, Universit\"at Heidelberg, Philosophenweg 16, 69120
  Heidelberg, Germany} 

\author{Mario Mitter}
  \affiliation{Physics Department, Brookhaven National Laboratory, Upton, NY 11973, USA} 

\author{Jordi Par\'is-L\'opez}
  \affiliation{Institute of Physics, NAWI Graz, University of Graz, Universit\"atsplatz 5, 8010 Graz, Austria} 
  \affiliation{Institut f\"ur Theoretische Physik, Universit\"at Heidelberg, Philosophenweg 16, 69120
  Heidelberg, Germany}

\author{Jan M. Pawlowski}
  \affiliation{Institut f\"ur Theoretische Physik, Universit\"at Heidelberg, Philosophenweg 16, 69120
    Heidelberg, Germany}
  \affiliation{ExtreMe Matter Institute EMMI, GSI, Planckstr. 1, D-64291 Darmstadt, Germany}
  
\author{Nicolas Wink}
  \affiliation{Institut f\"ur Theoretische Physik, Universit\"at Heidelberg, Philosophenweg 16, 69120
  Heidelberg, Germany}

\begin{abstract}
  We discuss an approach for accessing bound state properties, like
  mass and decay width, of a theory within the functional
  renormalisation group approach. An important cornerstone is the
  dynamical hadronization technique for resonant interaction
  channels. The general framework is exemplified and put to work
  within the two-flavour quark-meson model. This model provides a
  low-energy description of the dynamics of two-flavour QCD with quark
  and hadronic degrees of freedom. We compare explicitly the
  respective results for correlation functions and observables with
  first principle QCD results in a quantitative manner. This allows us to
  estimate the validity range of low energy effective models. We also
  present first results for pole masses and decay widths. Next steps
  involving real-time formulations of the functional renormalisation
  group are discussed.
\end{abstract}

\maketitle

\section{Introduction}
\label{sec:Introduction}
The efforts of determining bound state properties in a quantum field
theoretical approach date back to the seminal work of Bethe and
Salpeter \cite{Salpeter:1951zz,Salpeter:1951sz}. Despite considerable
progress in our understanding of bound states and their properties in
quantum field theories, the precise computation of their properties
and subsequently also the computation of spectra in general remains
one of the biggest challenges today.  Motivated by the immense
significance of reliable predictions of these quantities in
essentially all areas of physics there are quite a number of ongoing
investigations in this field. This applies in particular to QCD
because confinement makes only the composite states of quarks and
gluons, the hadrons, experimentally accessible.

The study of highly relativistic bound states has also been hampered
by the fact that almost all quantitative non-perturbative methods rely
on the Euclidean formulation of quantum field theory. This implies
that for determining bound state properties the results for
correlation functions have either to be continued to Minkowski space
or have to be extracted from potentially subleading exponential tails
of correlation functions. To this end we note that the existence of
stable bound states imply poles and of scattering states cuts in the
correlation functions for \textit{timelike} momenta. Unstable
particles and virtual states create in addition further poles in the
complex plane.  While the lowest excitations typically can be accessed
via reconstruction methods such techniques fail to provide trustable
results for the higher resonances. These higher-lying bound states
plainly require an exponentially enhanced precision of the imaginary
time data. Moreover, by definition, they lie beyond the radius of
convergence for Pad{\'e}-like analytic continuations of the imaginary
time results: Their masses are larger than the ground state mass, and
the latter provides the lowest-lying pole in exactly this channel.

In the present work we suggest a functional continuum approach to
bound state computations, which is also put in perspective to other
functional bound state approaches, for the impressive progress on
bound state and general low energy properties on the lattice we refer
the reader to, {\it e.g.}, \cite{Aoki:2016frl}.  Due to the bound
states poles in the correlation functions these states, and even
higher-lying resonances, can be accessed in continuum approaches via
the resonant frequency or momentum structures of higher-order
correlation functions of the fundamental degrees of freedom as, {\it
  e.g.}, quarks and gluons in QCD.  In case of gauge theories physical
states will hereby appear only in gauge-invariant channels. However,
an understanding of the non-perturbative properties of the elementary
correlation functions might necessitate the consideration of bound
states in unphysical channels, see, {\it e.g.},
\cite{Aguilar:2016ock,Aguilar:2017dco,Alkofer:2011pe}.

  The correlation functions that feature a pole at the bound state
  masses satisfy Dyson-Schwinger equations as all correlation
  functions of a theory, for respective reviews see, {\it e.g.},
  \cite{Alkofer:2000wg,Bashir:2012fs,Sanchis-Alepuz:2015tha}.  It has
  been exactly the achievement of Bethe and Salpeter to realise that
  employing a Laurent expansion of the respective correlation function
  around the bound state pole its non-linear and in general
  inhomogeneous Dyson-Schwinger equation (DSE) can be reduced to a
  linear and homogeneous equation, the Bethe-Salpeter equation (BSE),
  for a simpler quantity, the Bethe-Salpeter amplitude
  \cite{Salpeter:1951zz,Salpeter:1951sz}.  Hereby, the most
  challenging of the remaining complications is to find an
  approximation to the kernel of this equation which is on the one
  hand treatable and on the other hand keeps the most important
  symmetries of the underlying physics intact. The formulation and
  discussion of a systematic improvement of employed truncations for
  the kernel can be found in \cite{Sanchis-Alepuz:2015tha} and
  references therein. It has to be noted that the by far most used
  truncation in QCD and hadron physics, a generalised rainbow-ladder
  truncation, is for quite a number of meson and baryon channels quite
  successful, see, {\it e.g.}, the recent review
  \cite{Eichmann:2016yit}, but fails nevertheless for a large amount
  of hadron resonances as it can be inferred for example for a
  combined DSE-BSE approach for light mesons at the three-particle
  irreducible three-loop level \cite{Williams:2015cvx} or by recent
  investigations within the Functional Renormalisation Group (FRG)
  \cite{Mitter:2014wpa,Cyrol:2017ewj}.  It has to be emphasised,
  however, that keeping the most important symmetries at this level
  intact requires self-consistency of the treatment of a quite large
  amount of DSEs for the elementary QCD correlation functions together
  with the bound state BSEs. This then leads in such calculations to
  an overwhelming degree of complexity which effectively prevents the
  use of other available input data for QCD correlation functions,
  especially as the most sophisticated calculations imply a high
  degree of sensitivity of the bound state properties on details of
  the elementary QCD three-point functions.

  In the current work we discuss a unified approach to bound state
  properties based on the Functional Renormalisation Group (FRG).
  Hereby the correlation functions are not treated, at least in
  principle, as in the DSE-BSE approach by partly including the
  off-shell behaviour and partly restricting to on-shell properties.
  All these functions originate within the FRG approach from a common
  Effective Action. When employing dynamical hadronization
  \cite{Gies:2001nw,Pawlowski:2005xe,Floerchinger:2009uf,Floerchinger:2010da},
  the task of finding self-consistent truncation schemes is
  significantly facilitated. In particular, the challenge of staying
  within a class of truncations required to respect certain underlying
  symmetries can be resolved with comparatively less
  difficulties. Moreover, by now the FRG has matured enough to give a
  systematic access to bound state properties.

Here, we initiate this bound state program with a detailed study of
the quark-meson model in the light of the structures explained above.
We explain the natural embedding of this low-energy model as an
Effective Field Theory (EFT) of QCD as formulated in
\cite{Braun:2014ata,Mitter:2014wpa,Cyrol:2017ewj} within the fQCD
collaboration \cite{fQCD:2018-01}.  This includes the determination of
the EFT couplings directly in QCD from the QCD flows in
\cite{Mitter:2014wpa,Cyrol:2017ewj} leading to the QCD-assisted
quark-meson model.  Then we present results for the (Euclidean)
momentum dependence of correlation functions relevant for the access
to bound states.  As so far only the lightest states will be included,
an exploratory investigation into the Minkowski realm can still bee
done by means of a reconstruction of the real-time meson
propagators. This is notwithstanding the problems listed above for
reconstructions, and going beyond is our next step.

This paper is organized as follows: In Sect.~\ref{sec:bound_state_frg}
dynamical hadronization within the FRG is briefly revisited. The
adaption of this approach to QCD as well as the motivation for
employing the quark-meson model as a low-energy effective theory are
discussed in Sect.~\ref{sec:QCDfrg}. The truncation of the flow
equation for the quark-meson model is detailed in
Sect.~\ref{sec:qm_model}. Our results are then discussed in
Sect.~\ref{sec:Result}. In Sect.~\ref{sec:toQCD} the next steps which
are required to treat QCD are outlined. Our conclusions are presented
in Sect.~\ref{Conclusion}.  Technical details as well as a comparison
of results in different truncation levels are deferred to three
appendices.

\section{The FRG and bound states}
\label{sec:bound_state_frg}

The FRG approach to bound states as outlined in the present work is
based on two key ingredients, the calculation of real-time correlation
functions and the concept of Dynamical Hadronization, both of which
are briefly revisited in this section.

\subsection{Real-time FRG}
\label{sec:real_time_frg}
By now real-time versions of the FRG
have been developed that allow to access real-time
correlation functions, and, at the same time, making use of the rather
well-developed Euclidean correlation functions of the theories at
hand, and in particular QCD, for recent works see
\cite{Floerchinger:2011sc,Pawlowski:2015mia,Strodthoff:2016pxx,Pawlowski:2017gxj,
Kamikado:2013sia,Tripolt:2013jra,Tripolt:2014wra,Jung:2016yxl,Yokota:2016tip,
Wang:2017vis,Yokota:2017uzu, Bluhm:2018qkf}.

Here, our task is to extend this to dynamically generated bound state
properties.  To this end, regulators that preserve $SO(1,3)$ Lorentz
symmetry are of paramount importance, and we will focus our discussion
of the space-time symmetric regulators suggested in
\cite{Floerchinger:2011sc,Pawlowski:2015mia,Strodthoff:2016pxx,Pawlowski:2017gxj}.
Moreover, it is well-studied by now, that the rapid decay of the
regularised loops in frequency and momentum space is particularly
important in approximations for the full system that do not carry the
full frequency and momentum of the theory.  Alternatively, one can
study some questions about real time observables by applying
reconstruction methods. Combined with Euclidean FRG input this option
has been used in
\cite{Haas:2013hpa,Helmboldt:2014iya,Rose:2015bma,Tripolt:2016cya}. A
detailed analysis of the general complex structure of correlation
functions as well as the low and high frequency limits leads to
optimised reconstruction schemes, see \cite{Cyrol:2018xeq}. An
application of this novel reconstruction method to QCD spectral
functions can be also found in \cite{Cyrol:2018xeq}.

\subsection{Dynamical Hadronization}
\label{sec:intro_dyn_had}
Bound states and resonances in QCD and other theories manifest
themselves in resonant momentum channels in scattering amplitudes and
correlation functions. Typically, in an effective field theory
approach such a channel can be described as the exchange of an
effective field degree of freedom that carries the quantum numbers of
the resonant channel. This is the well-known Hubbard-Stratonovitch
(HS) transformation, originally introduced as an identity
transformation for a local four-point interaction of fermions done on
the level of a classical action.

The FRG allows to perform this identity transformation on the level of
the full quantum effective action, which avoids the well-known
double-counting problems of the HS transformation, if quantum
fluctuations are taken into account. This transformation, called
dynamical hadronization or more generally dynamical
condensation/bosonization has been introduced in \cite{Gies:2001nw}
and further developed in
\cite{Pawlowski:2005xe,Floerchinger:2009uf,Floerchinger:2010da}.
While originally introduced for the HS-type transformation for a
four-point function it is by now applicable to general and also
non-polynomial field operators \cite{Pawlowski:2005xe}. Its
applicability to the full effective action is intertwined with the
Wilsonian idea of integrating out fluctuations iteratively momentum
shell after momentum shell. As it is done as an exact identity
transformation at each RG step, it avoids any double counting
issue. This is the property that elevates it to an identity
transformation of the full quantum effective action.

One of the prominent advantages within such a formulation is the
following. Typically applications of functional methods
to strongly correlated systems such as QCD rely on systematic vertex
expansions in the absence of small parameters. The 'small' parameter
behind this systematics is the phase space suppression of the
contributions of diagrams with higher order vertices: These vertices
have no classical part and are generated by diagrams in the first
place, and hence possess for an asymptotically free theory 
 a rapid decay behaviour in momentum space which leads to,
after the angular integration,  a very efficient
suppression of the respective diagrams. This corresponds 
to a intuitive picture:  these $n$-order
vertices describe effectively the local interaction of $n$ particles which
is phase space suppressed. However, resonances in interaction channels do
not have this suppression if their regularised mass
is of the same order as the cutoff or below. 

Explicitly, for the case of QCD and hadron physics this implies
that  the occurrence of bound states or
resonances characterized by two or three valence quarks, {\it i.e.}, the 
mesons and baryons,  reduces 
scattering vertices of $(2n)$ or $(3n)$ quarks and anti-quarks to that
of $n$ mesons or baryons. This counter-acts significantly 
the phase space suppression of the relevant channels.
Accordingly one either goes to a higher order of the
vertex expansion in the fundamental fields in QCD or formulates QCD
also in terms of these additional effective degrees of
freedom. 

We close this discussion with two remarks on dynamical hadronization:
First, we emphasise again that dynamical hadronization does {\it not}
entail the reduction of QCD to a low energy EFT. It is only a
convenient and efficient reparametrisation of QCD in the dynamical low
energy degrees of freedom.  Second, even though the phase space
suppression is partially lifted in the presence of resonant
interactions, it is also the mass scale of these channels that decides 
about their relevance for quantum fluctuations.  In QCD these
resonances get strong at low, sub-GeV RG scales $k$.  Note that the
loop momenta in the FRG framework are restricted by the cutoff scale,
$p^2 \lesssim k^2$. Hence, all but the lowest lying resonances are
already decoupled when they are generated. Accordingly it is also
quantitatively sufficient to consider the dynamical hadronization of
the $\sigma$ mode and the pions $\vec \pi$.

Applications to QCD can be found in
\cite{Gies:2002hq,Braun:2014ata,Mitter:2014wpa,Cyrol:2017ewj}. In
particular, the references \cite{Mitter:2014wpa,Cyrol:2017ewj} contain 
an application of dynamical
hadronization on the quantitative level with full momentum
dependencies to QCD with the to date by far largest set of coupled
set of correlation functions in functional methods.

\section{Functional Renormalisation Group approach to QCD}
\label{sec:QCDfrg}

Here we give a brief introduction to the FRG-approach to QCD. More
details can be found in
\cite{Braun:2014ata,Mitter:2014wpa,Cyrol:2016tym,Cyrol:2017qkl,Cyrol:2017ewj},
a review on applications can be found in \cite{Pawlowski:2014aha}, 
for QCD-related FRG reviews see
\cite{Berges:2000ew,Pawlowski:2005xe,Gies:2006wv,Schaefer:2006sr,Braun:2011pp}.

\subsection{Functional Renormalisation Group}

The FRG is based on the Wilsonian idea of successively integrating
out quantum fluctuations restricted to a given momentum shell
$p^2 \approx k^2$ where $k$ is the running infrared cutoff scale of
the theory. The flow of the effective action is then governed by the 
Wetterich equation \cite{Wetterich:1992yh,Ellwanger:1993mw,Morris:1993qb}, 
\begin{align}
  \partial_t \Gamma_k[\Phi]=\frac{1}{2}\text{Tr}\,
  \frac{1}{\Gamma_k^{(2)}[\Phi]+R_k}\,
  \partial_tR_k\,,
\label{eq:FRG}
\end{align}
where $\partial_t =k\partial_k$, and the superfield $\Phi$ contains
all fields in the theory at hand. For example, in $N_f=2$ flavour
QCD with dynamical hadronization of scalar and pseudo-scalar
quark--anti-quark channels of the four-quark interaction we have
$\Phi=(A_\mu,c,\bar c,q,\bar q,\sigma,\vec \pi)$. This is the case
studied in \cite{Braun:2014ata,Mitter:2014wpa,Cyrol:2017ewj} relevant
for us. $\Gamma_k^{(2)}$ is the second field derivative of the
effective action, and
\begin{align}
\Gamma_k^{(n)}[\Phi]= \frac{\delta^n\Gamma[\Phi]}{\delta\Phi^n }\,.
\end{align} 
The trace in \labelcref{eq:FRG} sums over all species of fields,
internal indices and momenta, and $R_k(p^2)$ is a regulator function
that suppresses the propagation of the infrared modes of all fields.

In practice the flow equation \labelcref{eq:FRG} can only be solved within
approximations. As argued above, here we rely on a systematic vertex
expansion to QCD, for details we refer to
\cite{Mitter:2014wpa,Cyrol:2016tym,Cyrol:2017qkl,Cyrol:2017ewj}.
To that end we expand the
effective action in powers of the fields, 
\begin{align}
 \Gamma_k=\sum_{n\geq 1} \frac{1}{n!} \int_{p_1,\dots, p_n}\hspace{-.6cm} \Phi_1(p_1)\cdots
\Phi_n(p_n)\, \Gamma_k^{(n)}[p_1,\ldots, p_n]\,,
\label{eq:vertexexp}
\end{align}
and solve the flow equations for the $\Gamma_k^{(n)}$. This also
provides a systematic error estimate of the results with the
criterion of {\it apparent convergence}: The results apparently
converge if, when adding higher orders in the vertex expansion, no substantial changes occur. The system has then passed
a non-trivial self-consistency check. Note however, that this is but
one of the possible self-consistency checks strongly correlated
systems have to pass in the absence of a small expansion
parameter. 

 \subsection{Dynamical Hadronization}
 As already described in \Cref{sec:intro_dyn_had}, dynamical hadronization
 facilitates apparent convergence by means of restoring the canonical phase
 space suppression ordering of $n$-point functions.

 In the present case we use it for introducing an auxiliary scale dependent 
 mesonic field $\phi_k=(\sigma_k,\vec\pi_k)$ with 
 \begin{align}\label{eq:phik}
 \partial_t \sigma_k = \dot A_k \,\bar q q\,,\qquad 
 \partial_t \vec \pi_k = \dot A_k \,\bar q \,i  \gamma_5 \vec \tau q\,,
 \end{align}
 where the scale dependence $\dot A_k$ of this transformation can be chosen
 arbitrarily. This freedom is used to successively absorb the
 scalar--pseudo-scalar $u$-channel of the four-fermi scattering vertex,
 for more details see \Cref{app:DH}. Then, this channel
 vanishes identically in the four-fermi vertex.  Inserting \labelcref{eq:phik}
 in \labelcref{eq:FRG} we are led to a modified flow equation
 \begin{align}
 \left(\partial_t + \partial_t \phi_k[\Phi] \frac{\delta}{\delta\phi_k}\right) \Gamma_k[\Phi] =
 \frac{1}{2}\text{Tr}\,\frac{1}{\Gamma_k^{(2)}[\Phi]+R_k}\,\partial_tR_k\, .
 \end{align}
 The auxiliary field can then be interpreted as the resonance of that
 channel as it carries the physics and the same quantum
 numbers. Thereby dynamical hadronization allows for a convenient
 access to resonances and all associated bound state properties.

\subsection{QCD-assisted low energy effective theories} 
\label{subsec:QCDtoQM}

\begin{figure}[t]
	\begin{align*}
	\partial_t \Gamma_k[\Phi]
	=
	\frac{1}{2}\parbox[c]{0.078\textwidth}{\includegraphics[width=0.078\textwidth]{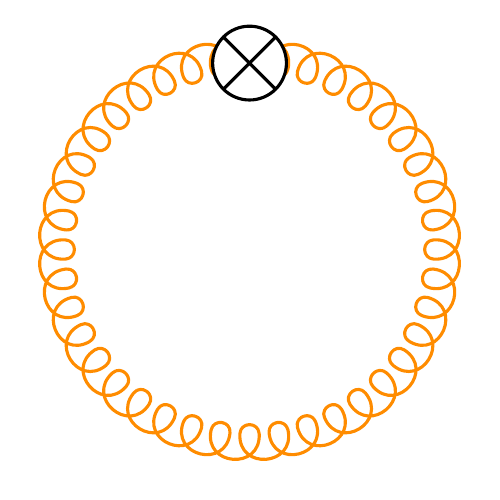}}
	-\parbox[c]{0.078\textwidth}{\includegraphics[width=0.078\textwidth]{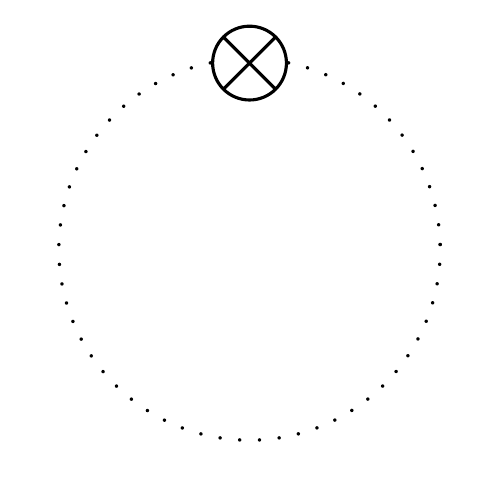}}
	- \parbox[c]{0.078\textwidth}{\includegraphics[width=0.078\textwidth]{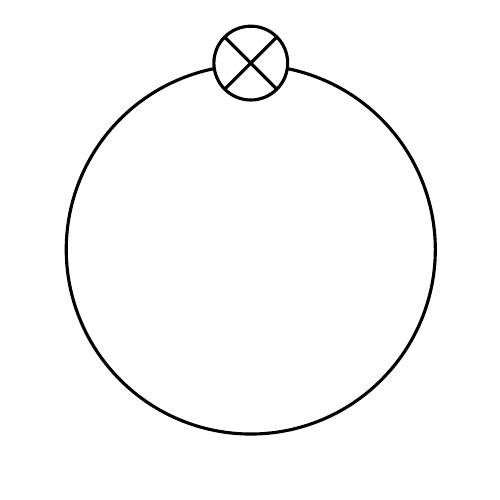}}
	+\frac{1}{2}\parbox[c]{0.078\textwidth}{\includegraphics[width=0.078\textwidth]{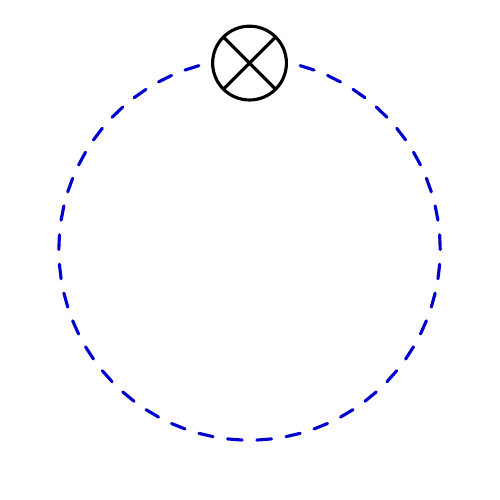}}
	\end{align*}
	\caption{Functional renormalisation group equation for QCD 
in which the $\sigma-\vec{\pi}$-channel is dynamically hadronized. 
The lines denote gluons, ghosts, quarks and mesons, respectively, and 
represent fully momentum and field dependent propagators. The cross denotes the 
regulator insertion $\partial_t R_k(p)$, leading to an effective UV cutoff for modes 
with $p^2\gtrsim k^2$.}
	\label{fig:fqcdrebosonised}
\end{figure}

The flow equation for QCD in the Landau gauge with dynamical
hadronization of the $\sigma$ and $\vec \pi$ four-quark channels is
depicted in \Cref{fig:fqcdrebosonised}. When lowering the cutoff scale
$k$ successively, the single loops in \Cref{fig:fqcdrebosonised} are
suppressed if the cutoff scale drops below the mass gap of the
respective propagators. This has very important consequences.

First of all, the meson loops do not contribute at large cutoff scales: 
The effective mesonic mass scale drops rapidly for large cutoffs (and
momenta) because the respective momentum channels in the four-quark
interaction decay rapidly. This in turn leads to an increased importance
of  the mesonic loops at low cutoff scales. This is in agreement with the
 pion mass being  the smallest mass scale. 

\begin{figure}[t]
	\begin{align*}
	\partial_t \Gamma_k[\Phi_\tinytext{EFT}]
	=
	- \parbox[c]{0.078\textwidth}{\includegraphics[width=0.078\textwidth]{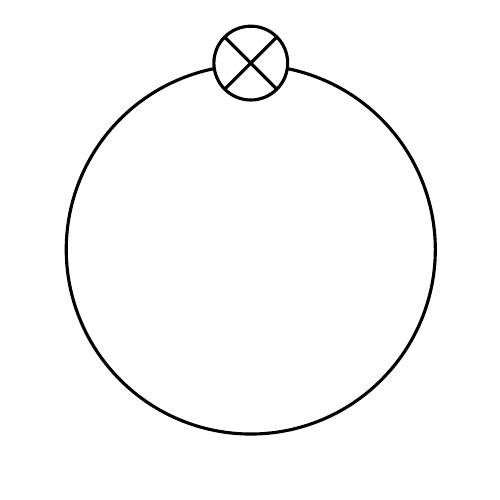}}
	+\frac{1}{2}\parbox[c]{0.078\textwidth}{\includegraphics[width=0.078\textwidth]{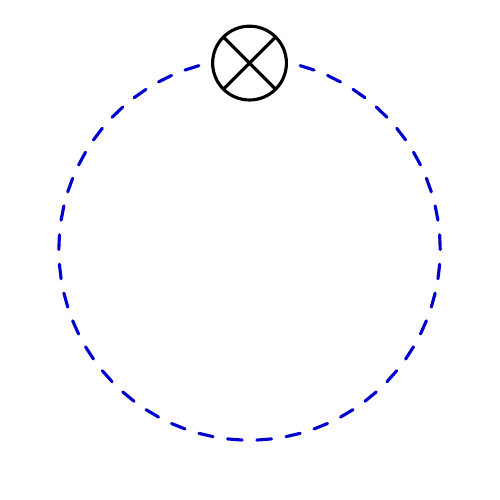}}
	\end{align*}
	\caption{Functional renormalisation group equation for the two-flavour
		quark-meson model, for the notation {\it cf.} \Cref{fig:fqcdrebosonised}.}
	\label{fig:frgQM}
\end{figure}

Second, the gluon effectively decouples below cutoff scales of
$k\lesssim 1\,$GeV as the gluon is gapped in the Landau gauge with a
mass gap of almost one GeV. Hence, for these scales the gluon loop can
be dropped in \Cref{fig:fqcdrebosonised} for
$k\lesssim \unit[1]{GeV}$. Moreover, to leading order the ghost fields
only couple to the matter part of QCD via the gluon. Accordingly, they
effectively decouple from the matter sector of QCD as well at the same
decoupling scale as the gluon. This leaves us with a quark-meson flow
equation at low cutoff scales which is depicted in \Cref{fig:frgQM}.

In summary, this leaves us with a QCD-assisted low energy effective
theory for QCD. Its natural ultraviolet cutoff $\Lambda$ is at almost
$\unit[1]{GeV}$, and the 'classical' action
$S_{\text{\tiny{EFT}},\Lambda}$ of this EFT is the full effective
action of QCD at this scale, evaluated on the equations of motion for
the gauge field and the ghosts: $A_\mu=0\,,\,c\,=0=\bar c$ and thus
\begin{align}\label{eq:S-EFT}
S_{\text{\tiny{EFT}},\Lambda}[q,\bar q, \phi] = \Gamma_{\text{\tiny{QCD}},\Lambda}[
\Phi_{\text{\tiny{EFT}}}]\,,
\end{align} 
with the superfield 
\begin{align}
\Phi_{\text{\tiny{EFT}}}=(0,0,0,q,\bar q,\phi)\,,\qquad \phi=(\sigma,\vec \pi)\,.
\end{align}
This concludes the discussion of the QCD-embedding of the low energy
EFT under investigation here, additional discussions of this issue can
e.g. be found in \cite{Papp:1999he,Braun:2014ata,Springer:2016cji,Eser:2018jqo}.

\section{Quark-Meson model}
\label{sec:qm_model}
Utilizing the previous discussion we can write down the leading terms
of the QCD effective action at a scale where the ghost and gluons are
to a large degree already decoupled. Then, the QCD effective action
with dynamical hadronisation as described in the previous section
reduces to a low energy effective theory with quarks and mesons, the
quark-meson (QM) model. Bound state considerations in the QM model
were amongst the first application of the FRG, see e.g.\
\cite{Jungnickel:1995fp}, for early reviews see
\cite{Jungnickel:1998fj,Berges:2000ew,Schaefer:2006sr}.

\begin{align}
\Gamma_{k} = \int_{x} \Bigg\{
&Z_{q,k} \overline{q} \slashed{\partial}  q
+\lambda_{k} \left[ \frac{1}{2 N_f} \left( \overline{q} q \right)^2
- \left( \overline{q} \gamma_5 \vec{\tau} q \right)^2\right] \nonumber \\[1ex] 
+& \frac{1}{2} Z_{\sigma,k} (\partial_\mu \sigma)^2
+ \frac{1}{2}Z_{\pi,k}(\partial_\mu \vec{\pi})^2  
+V_k -c \sigma\nonumber
 \\[1ex]
+& h_{k}  \overline{q} \left( \imag \gamma_{5} \vec{\tau} \cdot \vec{\pi} 
+ \frac{1}{\sqrt{2 N_f}} \sigma \right) q 
\Bigg\} \, .
   \label{eq:qm_model} \end{align}
 The effective action \labelcref{eq:qm_model} consists of three parts: the
 fer\-mionic and the bosonic ones as well as a Yukawa interaction 
 in between both of them. The latter two, {\it i.e.}, the second and 
 third line in the effective action \labelcref{eq:qm_model}, result from dynamical
 hadronization.  As for the first line, 
 the quarks'  kinetic term appears already in the
 classical action of QCD.  On the other hand, the four-quark
 interaction is induced by the quark-gluon interaction and is thus
 created dynamically during the evolution of the scale
 dependent effective action. Herein, we only consider the scalar--pseudo-scalar 
 $u$-channel of the four-fermi vertex, which is by far the most dominant one (see the 
 discussion above).  Consequently, the bosonic part contains the propagators 
 for the sigma meson and the pions. These terms as well as the Yukawa interaction are
 already obtained by a single Hubbard–Stratonovich transformation
 (see, {\it e.g.}, Chapter 3 of \cite{Alkofer:1995mv}).  Higher-order
 terms induce self-interactions between the mesons which are
 described by the effective potential $V_k$. Since isospin symmetry remains 
 unbroken it is a functional of the $O(4)$-symmetric combination
 $\rho = \frac{1}{2}\left(\sigma^2 + \vec{\pi}^2\right)$.  A non-vanishing
 expectation value of $\rho$, $\langle\rho\rangle \not= 0$, 
 signals condensation of the sigma meson
 and thus spontaneous symmetry breaking. Current quark masses lead
 via an appropriate shift of the $\sigma$ field to the last term 
 in the second line of the effective action \labelcref{eq:qm_model}. It 
 explicitly breaks $O(4)$-symmetry and effectively alters the
 expectation value of the sigma meson by tilting the effective potential, 
 which in turn results in a 
 finite mass for the Goldstone modes, {\it i.e.}, the pions.  
 To summarize, the ansatz \labelcref{eq:qm_model} for the scale-dependent 
 effective action captures within a reasonable approximation
 the leading behaviour of two-flavour QCD at
 low momenta, a statement tested explicitly in
 \Cref{sec:result_euclidean}.

In the effective action \labelcref{eq:qm_model} the wave-function
renormalization functions, $Z_{\sigma,k}(p), Z_{\vec{\pi},k}(p)$ and
$Z_{q,k}(p)$, the Yukawa coupling $h_k(p_1,p_2)$ and the four-fermi
coupling $\lambda_k(p_1,p_2,p_3)$ are momentum-dependent quantities.
Please note that we employ a notation in which momentum conservation
is exploited and the three-point vertex $h_k$ is a function of two
independent momenta, the four-point vertex $\lambda_k$ of three
momenta. To this end, in the Yukawa coupling the quark momenta are
singled out As already mentioned, the successive Hubbard–Stratonovich
transformations in the Dynamical Hadronization are then used to cancel
the flow of the four-fermi coupling, {\it i.e.}, one implements
\begin{align}
\partial_t \lambda_k(p,p,-p) = 0
\, .
\end{align}
which yields an additional contribution to the flow of the Yukawa
coupling.

In the following we differentiate between bare quantities and
renormalized quantities which are then denoted by a bar.
Renormalization conditions are imposed by requiring
\begin{align}
\bar{Z}_{\vec{\pi},k}(p=0) = 1 \qquad \text{and} \qquad \bar{Z}_{q,k}(p=0) = 1
\, .
\end{align}
As a result all dressed quantities are RG invariant, while the bare
ones are not.

Further details about the effective action \labelcref{eq:qm_model},
the solution of the respective flow equation ({{\it cf.}
  \labelcref{eq:FRG}), and related technical details can be found in
\Cref{app:add_details}.

Exploiting the freedom that the four-fermi coupling
$\lambda_k(p_1,p_2,p_3) $ can be set to zero at the ultraviolet (UV)
cutoff, {\it i.e.}, for $k=\Lambda_\text{\tiny{UV}}$, the four-fermi
coupling is absent in all flow equations to be solved in the
following.  Previous studies in QCD, \cite{Cyrol:2017ewj}, have shown
that the full momentum dependence of the resulting Yukawa vertex can
be well approximated by a function of only one momentum upon the
substitution rule
\begin{align}
h_k(p_1,p_2) \to h_k\left( \sqrt{\frac 1 4 (p_1-p_2)^2 + (p_1+p_2)^2} \right).
\end{align}
This will be employed in most of the results presented in the next
section.

It is also important to note that the quark mass function and the
Yukawa coupling are related via the $\sigma $ condensate,
\begin{align}
m_{q,k}(p) = \frac{ h_k(p,-p)}{\sqrt{2 N_f}} \langle\sigma\rangle
\, .
\end{align}
The pion decay constant $f_\pi$ will in the following be calculated
from the $\sigma $ condensate based on an approximation provided by
the Gell-Mann--Oakes--Renner relation. The accuracy of this strongly
simplified way to determine the pion decay constant is sufficient for
the purpose of the exploratory investigation presented here.

\begin{figure*}[t!]
	\includegraphics[width=\textwidth]{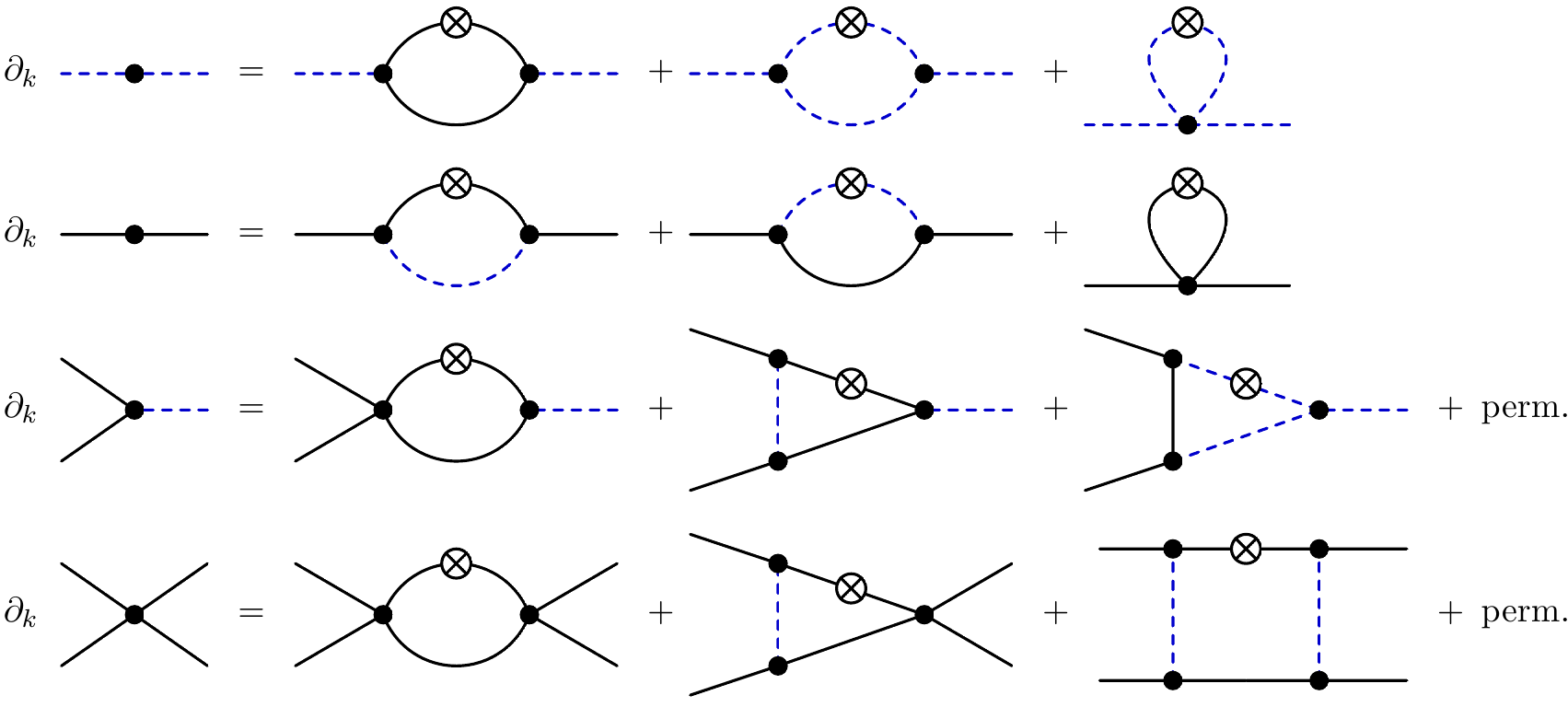}
	\caption{Flow equations for the propagators, Yukawa coupling
          and four-fermi interactions.}
	\label{fig:FlowEq}
\end{figure*}

The initial values for the flow are chosen at the UV cutoff such that
physical values for observables are achieved in the infrared $k\to 0$.
This includes setting the pion decay constant to
$f_\pi=\unit[93]{MeV}$, and the curvature masses to
$\overline{m}_{\pi}=\unit[138.7]{MeV}$,
$\overline{m}_\sigma=\unit[500]{MeV}$, and
$\overline{m}_{q}=\unit[297]{MeV}$.  The specific details on the
precise definitions, respectively, the approximations employed for
determining the masses are given in the next section.  The remaining
flow equations can be derived from \Cref{fig:frgQM}. This results in
the set of equations depicted in \Cref{fig:FlowEq}. The explicit
expressions for the flow equations are deferred to
\Cref{app:flow_equations}.

\section{Results and discussion}
\label{sec:Result}
We start in \Cref{sec:result_euclidean} with a discussion of the
results for Euclidean momenta. The extension of the results to the
real time domain for the extraction of bound state properties is
discussed in \Cref{sec:result_minkowski}.

\subsection{Solution of the flow equation to Euclidean momenta}
\label{sec:result_euclidean}
The main results of this section are the momentum dependencies of the
propagators as well as the Yukawa coupling. In addition to the
physical parameters outlined in \Cref{sec:qm_model} we are considering
an additional parameter set where the IR has been fixed to the values
of a first principle calculation of QCD correlation functions with the
FRG presented in \cite{Cyrol:2017ewj}.  This comparison has the
advantage of testing the validity of low energy effective descriptions
of bound states in QCD.

\begin{figure*}[t]
	\includegraphics{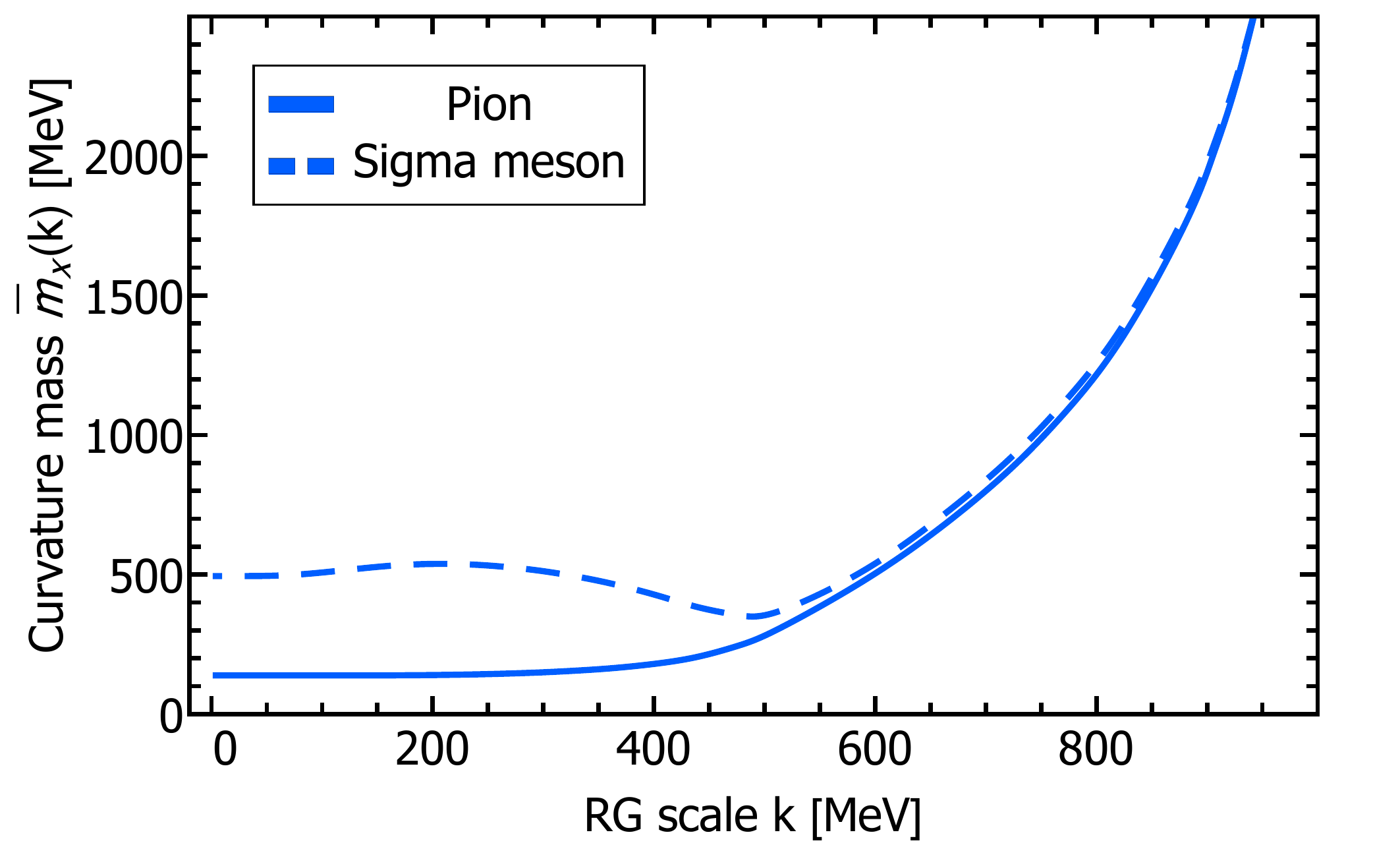}\hfill
	\includegraphics{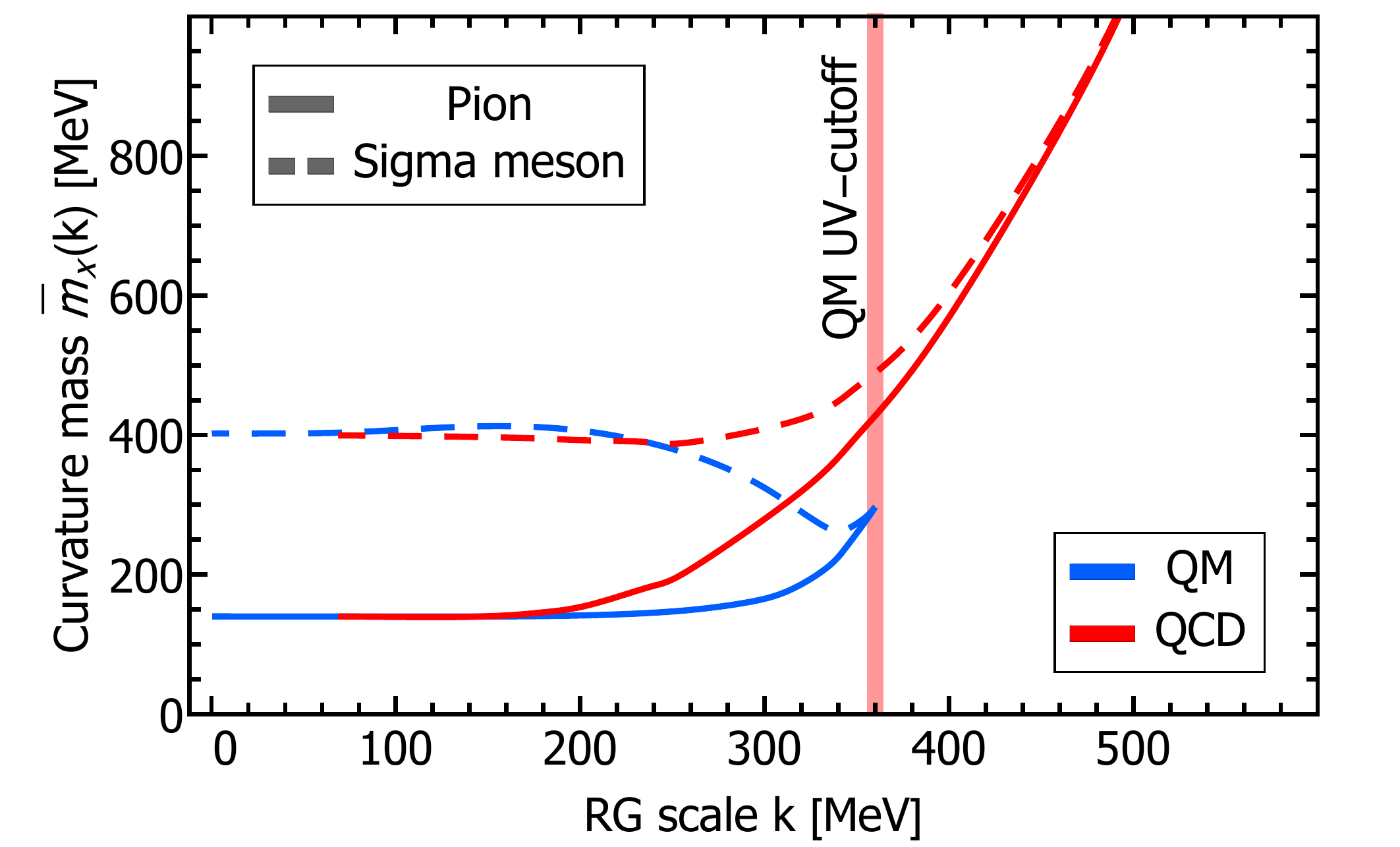}
	\caption{The flow of the curvature masses for the mesons in
          the quark-meson model (left panel).  The flow is started in
          the chirally symmetric regime.  During the flow chiral
          symmetry is broken dynamically and thus the masses of pion
          and sigma run apart in the Quark-Meson Model. In the right
          panel the results are compared when the flow of the
          quark-meson-model is QCD-assisted, and compared to those in
          full QCD from \cite{Cyrol:2017ewj}.}
	\label{fig:curvature_mass_flow}
\end{figure*}

\begin{figure*}[t]
	\includegraphics{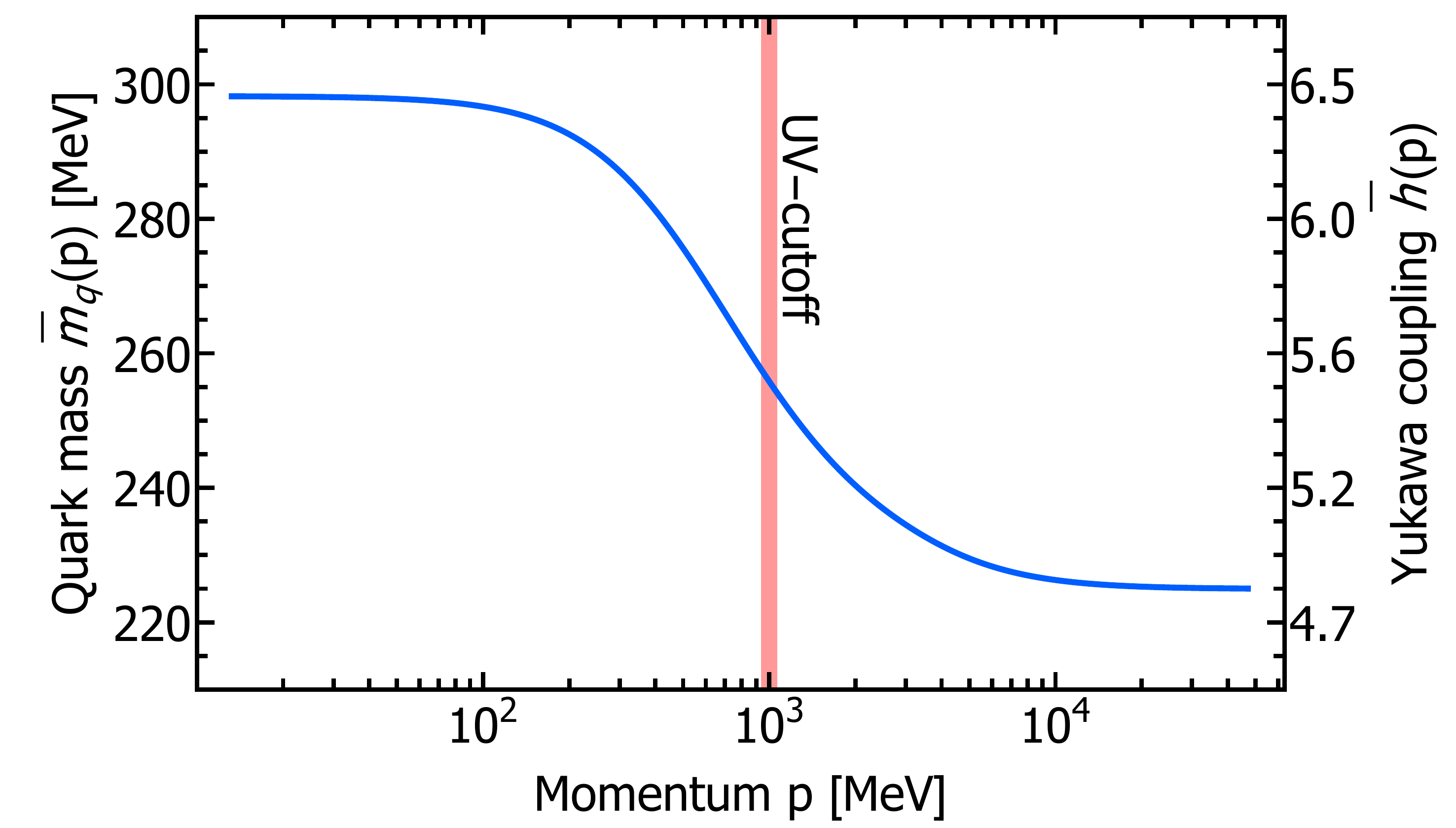}\hfill
	\includegraphics{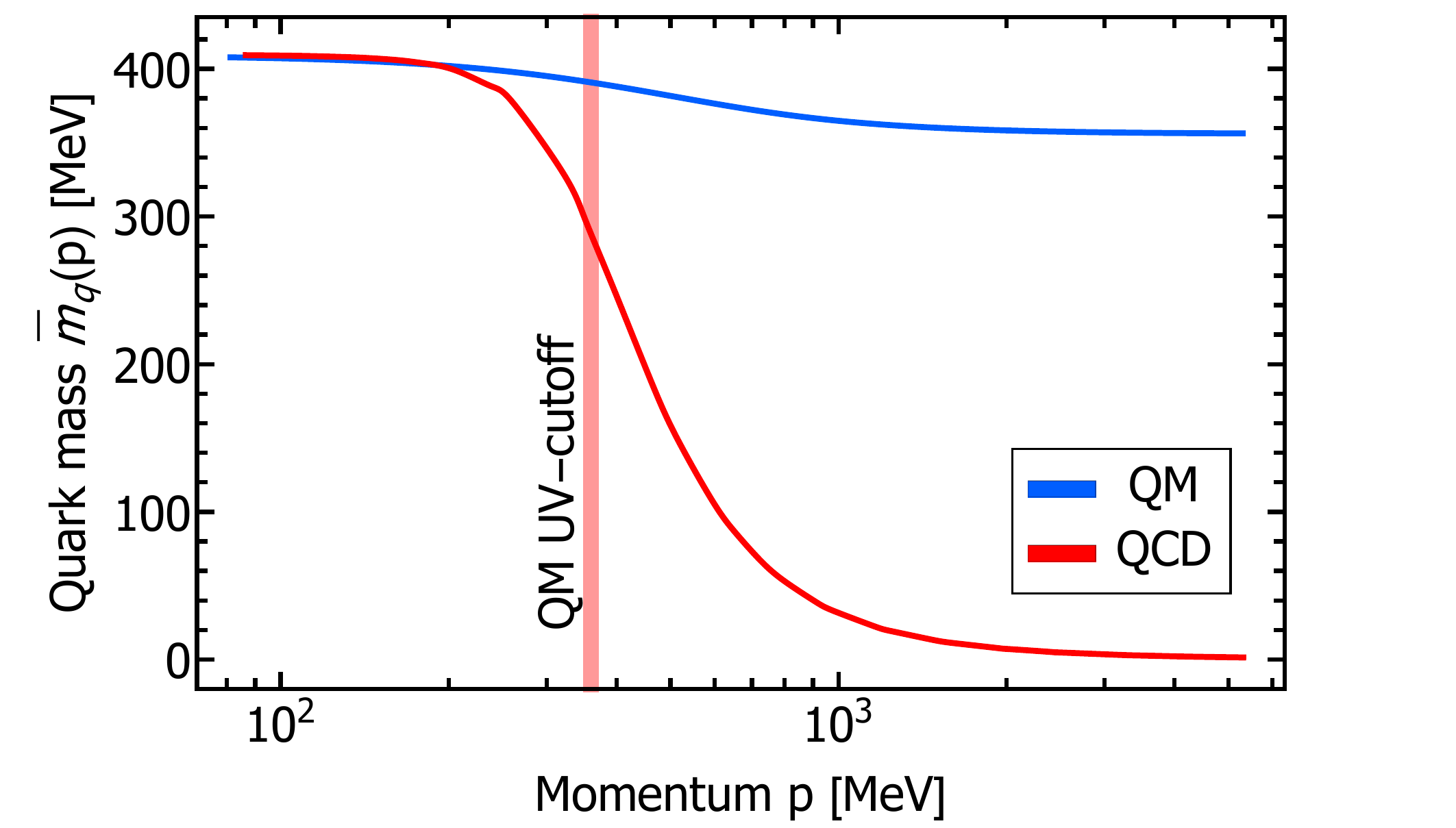}
	\caption{The momentum-dependent quark mass function in the
          quark-meson model (left panel). The right panel shows a
          comparison of the same quantity between the QCD-assisted
          quark-meson model and QCD.}
	\label{fig:quark_mass_flow}
\end{figure*}

\begin{figure*}[t]
	\includegraphics{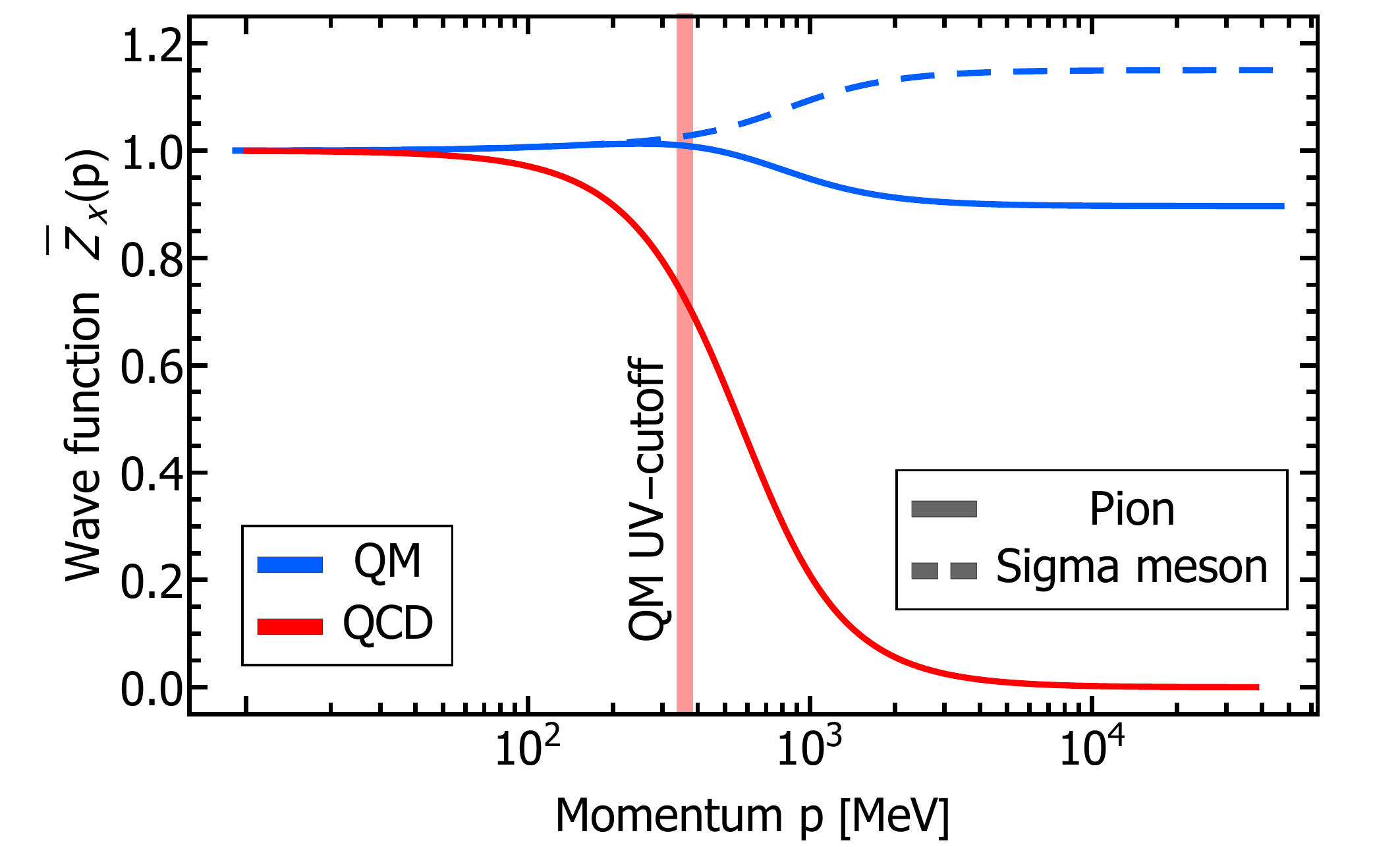}\hfill
	\includegraphics{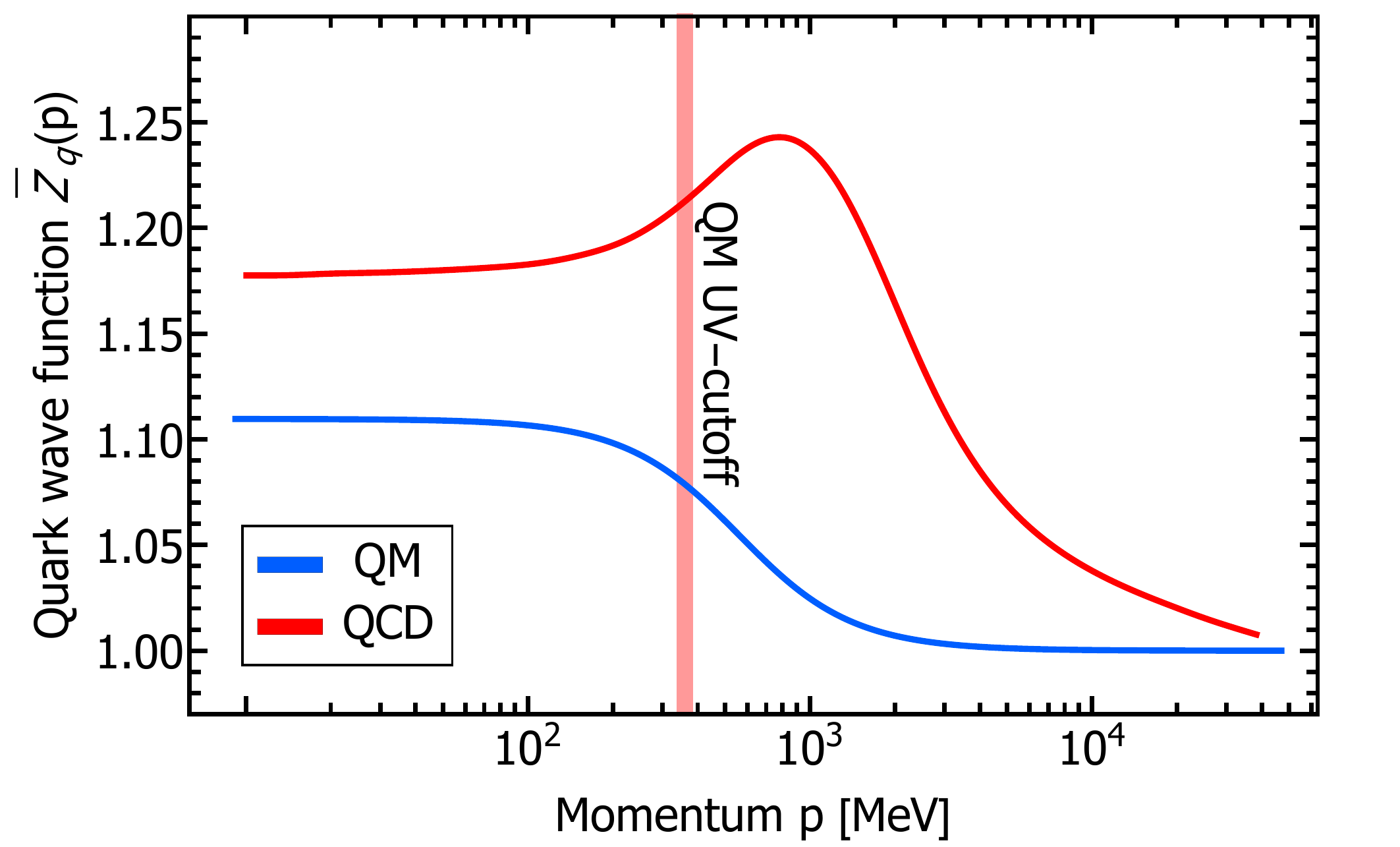}
	\caption{Comparison between the QCD-assisted QM model and QCD
          for the momentum-dependent wave function renormalizations
          for the mesons (left panel) and quarks (right panel).}
	\label{fig:wavefunctions}
\end{figure*}

\begin{figure}[t]
	\includegraphics[width=0.5\textwidth]{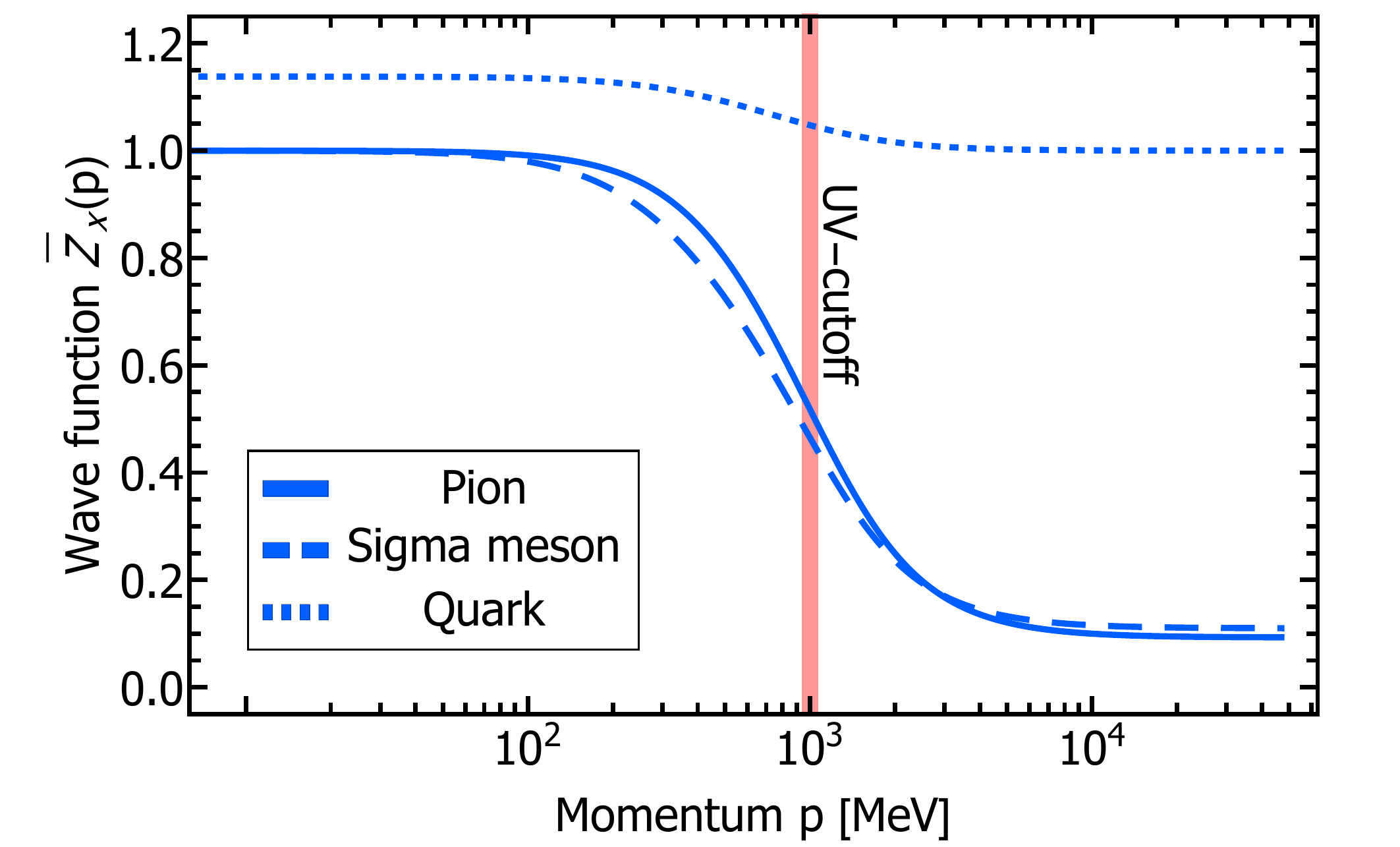}
	\caption{The momentum-dependent wave-function renormalization
          of the mesons and quarks in the quark-meson model.}
	\label{fig:wavefunction_QM}
\end{figure}

The first quantity of interest from an RG point of view is the flow of
the curvature masses evaluated at the flowing minimum of the effective
potential, depicted in \Cref{fig:curvature_mass_flow}.  The flow of
the curvature masses is a key ingredient in all diagrams because it
sets the dominant scale in integrands.  In the left panel the result
for the quark-meson model with physical IR values is shown.  The
result is qualitatively similar to previous calculations in the
quark-meson model (QM), see, {\it e.g.},
\cite{Pawlowski:2014zaa,Rennecke:2016tkm}.

Since we are mostly interested in the \ description of bound states,
{\it i.e.}, low-energy effective degrees of freedom, the comparison to
the full QCD calculation is also of particular interest, since the
curvature mass determines mainly the scale in the resulting
correlation functions.  The comparison is depicted in the right panel
of \Cref{fig:curvature_mass_flow} and shows that they agree quite well
for small and moderate scales.  This demonstrate that the QM model
indeed describes the dynamics of low energy QCD correctly.  However,
in order to achieve the same IR values in the QM setting as in the QCD
calculation the UV-cutoff of the theory has to be lowered to
$\sim \unit[360]{MeV}$, see also \cite{Rennecke:2015}. With a higher
UV- cutoff it is not possible to decrease the sigma mass further while
keeping $f_\pi$, $m_\pi$ and $m_q$ fixed. This is most likely linked
to the triviality of the O(N) model, which shrinks the values of
achievable IR values for a given cutoff when increasing the
truncation, which has already been observed in \cite{Marko:2016wtw,
  Pawlowski:2017gxj}.

The most significant differences between the QM model and the full QCD
calculation is the decoupling of the mesonic masses from the system,
which can be seen from comparing both subfigures in
\Cref{fig:curvature_mass_flow}: First, the mesonic degrees of freedom
decouple substantially faster in full QCD. This is also immediately
visible when looking at the quark mass in \Cref{fig:quark_mass_flow}
or at the wave function renormalizations in figures
\Cref{fig:wavefunctions} and
\Cref{fig:wavefunction_QM}. Second, the mass function of the
  scalar $\sigma$-channel in QCD does not show the dip towards smaller
  masses in the region where chiral symmetry is restored, which is
  also related to the steeper rise of the mass function in QCD. This
  feature has potentially important consequences for the details of
  chiral symmetry restauration at finite temperature that deserve
  further investigation. There, the low UV cutoff scale additionally
  results in a relatively low highest temperature that can be
  considered due to the the thermal range of these models, see
  \cite{Helmboldt:2014iya}. More generally, the necessary modification
  of the initial effective action for large external scales such as
  temperature, chemical potential, external background (chromo-)
  magnetic and electric fields has to be considered, for a detailed
  discussion see \cite{Braun:2018svj}. Again the systematic inclusion
  of these modifications is facilitated within a QCD assisted low
  energy theory.

In all plots the red vertical line denote
the UV cutoff used in our calculation. The slow decay for large
momenta is not very surprising in the quark-meson model as there is no
other scale involved which could potentially suppress the mesonic
degrees of freedom. Nevertheless, it is one of the most prominent
differences to the full QCD calculation, resulting in a rather small
range of momenta, where the quark-meson model properly describes all
dynamics of full QCD. However, this does not imply that the
description at larger momenta is bad, merely that the question depends
strongly on the observable at hand.

\subsection{Continuation to timelike momenta}
\label{sec:result_minkowski}
The calculation of observables requires in general the knowledge of
correlation functions in Minkowski spacetime. In functional methods
this reduces to essentially two distinct options, the direct
calculation via analytically continued equations or the numerical
analytic continuation.  While the former way is preferable, its
application to the quark-meson model is postponed to future work. Here
we resort to the latter option, which is for our purposes, {\it i.e.},
the extraction of the pole masses of the lowest-lying bound states,
quite accurate.  In addition, numerical analytic continuation is
easily possible for this case because the analytic structures of
interest are the poles closest to the origin in their respective
analytically continued retarded Greens functions, see
\cite{Cyrol:2018xeq} as well as references therein for a respective
detailed discussion. Moreover, it has been already shown in
\cite{Helmboldt:2014iya} that the momentum-dependence of the mesonic
propagator is rather mild which also facilitates the reconstruction of
the lowest-lying pole. In summary, this allows us to use a
\textit{Pad{\'e}} based approach, {\it i.e.}, a rational
interpolation. More specifically, we use the Schlessinger's point
method \cite{Schlessinger1968} to compute the interpolation from a
subset of data points.

\begin{table}[h]
\begin{ruledtabular}
\begin{tabular}{@{}*{7}{c}}
  Particle& Curvature  mass  & Pole  mass & Decay-width \\[1ex]
  Pion& 138.7	 & 137.4 $\,\pm\,$ 0.5 & 0.5 $\,\pm\,$ 0.5\\[1ex]
  Sigma & 494.5 & 320 $\,\pm\,$ 25 & 36 $\,\pm\,$ 5 \\
\end{tabular}
\end{ruledtabular}
\caption{Curvature mass, pole mass and decay width, as extracted from
  the full momentum dependence of the respective propagators. A
  significant deviation between the pole and curvature mass of the
  sigma meson is observed. All values are given in MeV.}
 \label{tbl:pole_masses}
\end{table}

\begin{figure}[t]
	\includegraphics[width=0.5\textwidth]{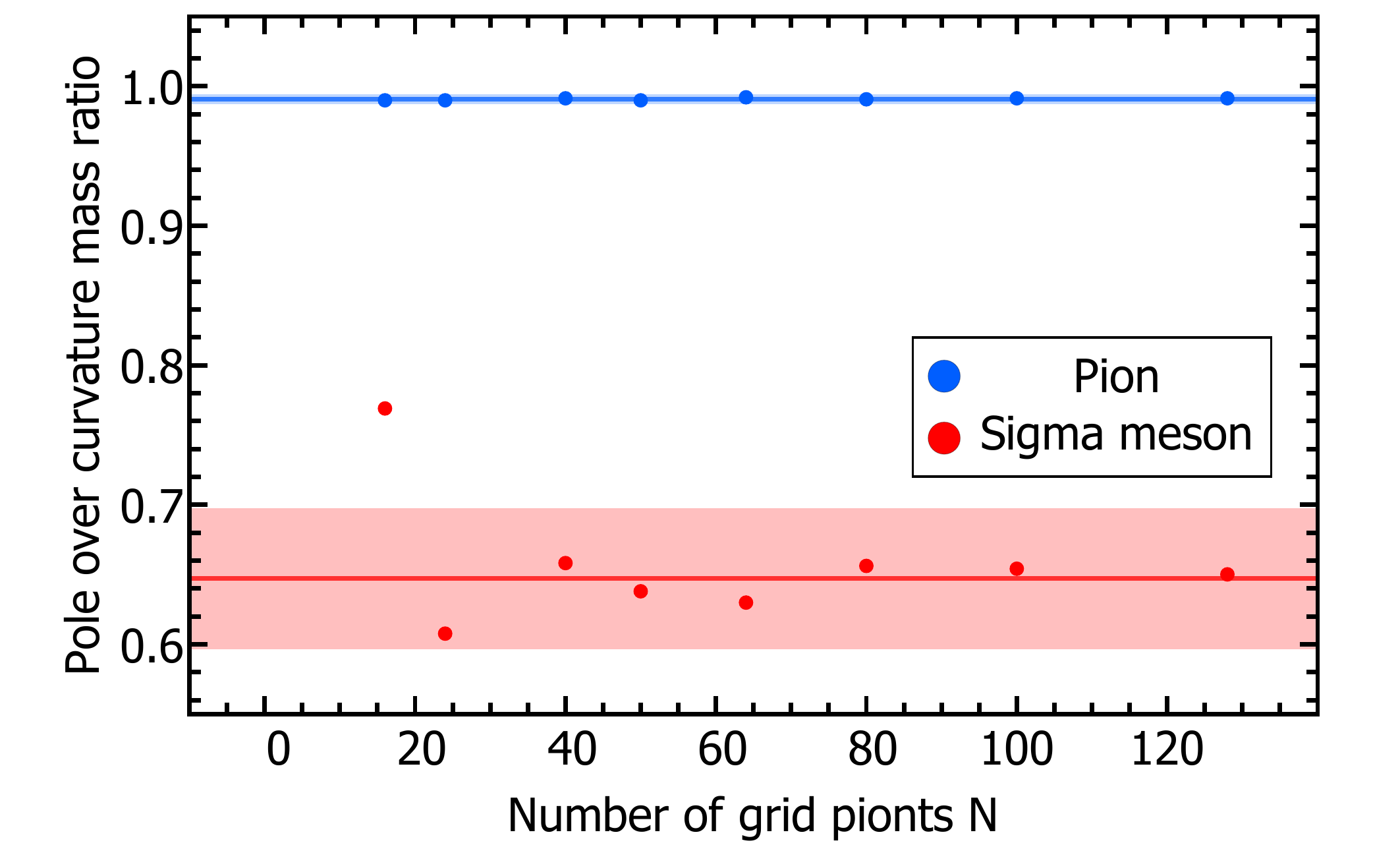}
	\caption{The ratios of the meson pole masses $m_{\text{pole}}$
          and curvature masses $m_{\text{cur}}$ are shown. The ratio
          is stable for the pion and converges slowly, but eventually,
          for the sigma as a function of grid points $N$.}
	\label{fig:Pade_Pol}
\end{figure}

The pion is the lightest degree of freedom in the quark-meson model
and thus stable.  For the sigma meson it is found in the present
calculation that its pole mass slightly exceeds the two pion decay
threshold. Therefore, we expect the mass to be dominated by a pole
close to the timelike axis on the second Riemann sheet of the retarded
propagator and therefore accessible within this framework to good
accuracy. The additional numerical error of the reconstruction is
checked by computations with different numbers of grid points.  The
resulting masses and widths are collected in \Cref{tbl:pole_masses}
and displayed relative to the curvature masses in
\Cref{fig:Pade_Pol}. As expected from analytic arguments the pole mass
of the pion agrees well with its renormalized curvature mass
\cite{Helmboldt:2014iya, Pawlowski:2017gxj}, and its decay width is
zero within numerical uncertainties. As outlined above the sigma meson
features in our calculation within the quark-meson model only a small
decay width, $\Gamma/M\ll 1$, validating our reconstruction
approach. Furthermore, the relative difference between the pole and
curvature mass is significant with $\sim35\%$.  Such an order of
magnitude agrees qualitatively with previous studies
\cite{Pawlowski:2017gxj} in which the momentum dependence of the
propagators was not fed back into the equations.  This has obvious
consequences for low energy effective theories of QCD where the mass
of the sigma meson is often used to fix the free parameters.
\begin{figure}[t]
	\begin{subfigure}{.25\textwidth}
		\centering
		\includegraphics[width=.8\linewidth]{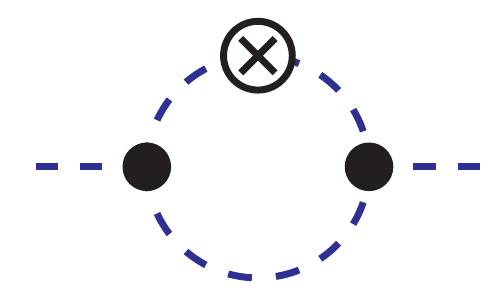}
		\caption{Mesonic contribution}
		\label{fig:meson_polarisation_with_meson}
	\end{subfigure}%
	\begin{subfigure}{.25\textwidth}
		\centering
		\includegraphics[width=.8\linewidth]{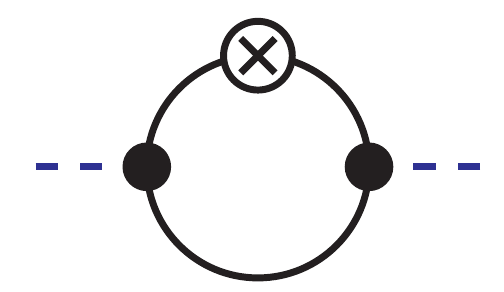}
		\caption{Quark contribution}
		\label{fig:meson_polarisation_with_quark}
	\end{subfigure}
	\caption{The two dominant diagrams contributing to the wave functions of the mesons in QCD (up to permutations of the regulator).}
	\label{fig:dominant_diags_polarisation}
\end{figure}

\section{Systematic improvements towards QCD}
\label{sec:toQCD}

The discussion in \Cref{sec:Result} enables us to systematically
improve the current setting towards QCD on a quantitative
level. Considering the structure of \Cref{fig:frgQM}, or directly in
the corresponding equations in \Cref{app:flow_equations}}, it is
immediately clear that only three quantities that are not directly
derived from the effective potential enter the equations:
$\bar{Z}_{\phi}(p)$, $\bar{Z}_q(p)$ and $\bar{h}(p,r)$. Please note
that we do not need to differentiate between $\bar{Z}_\pi(p)$ and
$\bar{Z}_\sigma(p)$ as their difference is negligible for all points
discussed here, {\it cf.} \Cref{fig:wavefunction_QM}.

While the difference in the quark wave function $\bar{Z}_q(p)$ between
the quark-meson model and QCD might appear large in
\Cref{fig:wavefunctions}, the values stay practically at unity for all
scales and the bump in QCD at the scale of the gluon mass gap
corresponds only to a sub-leading quantitative correction. Therefore,
we can concentrate our discussion on $\bar{Z}_{\phi}(p)$ and
$\bar{h}(p,r)$.

\begin{figure}[t]
	\begin{subfigure}{.15\textwidth}
	\centering
	\includegraphics[width=.8\linewidth]{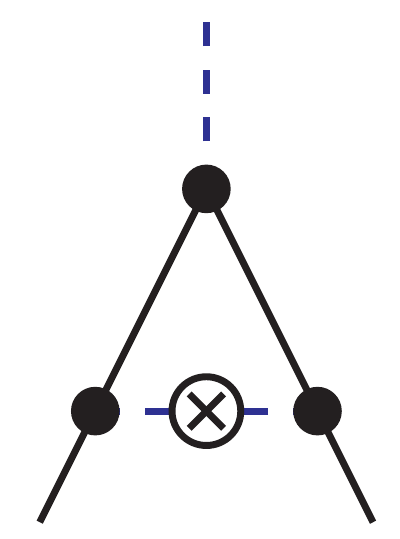}
	\caption{Quark-Meson vertex}
	\label{fig:yukawa_triangle_with_meson}
	\end{subfigure}%
	\begin{subfigure}{.15\textwidth}
		\centering
		\includegraphics[width=.8\linewidth]{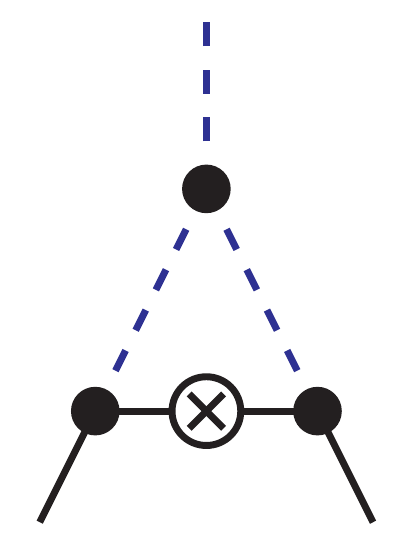}
		\caption{Quark-Meson vertex}
		\label{fig:yukawa_triangle_with_meson_2}
	\end{subfigure}%
	\begin{subfigure}{.15\textwidth}
		\centering
		\includegraphics[width=.8\linewidth]{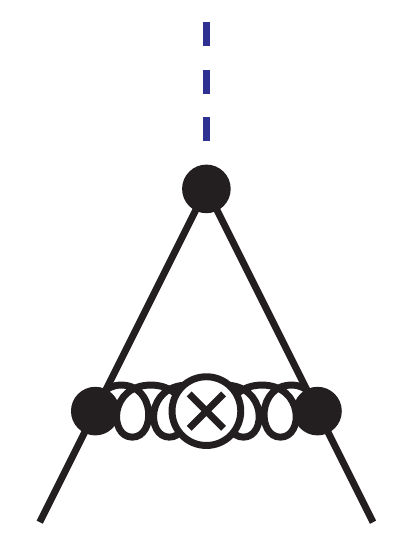}
		\caption{Quark-Gluon vertex}
		\label{fig:yukawa_triangle_with_gluon}
	\end{subfigure}
	\caption{The three dominant diagrams contributing to the
          Yukawa coupling in QCD (up to permutations of the
          regulator).}
	\label{fig:dominant_diags_yukawa}
\end{figure}

Turning to the pion and sigma meson wave functions, significant
differences between the quark-meson and the QCD result are observed,
see \Cref{fig:wavefunction_QM}. However, the leading diagrams that
generate the momentum dependence are the same in both cases and shown
in \Cref{fig:dominant_diags_polarisation}. As a result, the main
difference between the quark-meson and QCD results must be generated
by $\bar{h}(p,r)$.  In order to confirm this, we can apply the same
reasoning to the leading diagrams of the Yukawa coupling in QCD shown
in \Cref{fig:dominant_diags_yukawa}.  The first two diagrams,
\Cref{fig:yukawa_triangle_with_meson} and
\Cref{fig:yukawa_triangle_with_meson_2}, are again self-consistently
contained in the quark-meson model. However, the diagram containing a
gluon, \Cref{fig:yukawa_triangle_with_gluon}, is not. It is not
possible to include this diagram without the knowledge of the gluon
propagator and the quark-gluon vertex building up the one gluon
exchange coupling, but QCD-assisted models can be constructed with
this input. We conclude from this analysis that the quark-meson model
can be systematically improved towards QCD by either including the QCD
Yukawa coupling as external input or the gluon exchange coupling in
\Cref{fig:yukawa_triangle_with_gluon}. This underlines the strength of
the current setting of systematically improvable QCD-assisted
low-energy effective theories.

\section{Conclusion}
\label{Conclusion}
In this work we have outlined an approach to the calculation of bound
states within the Functional Renormalisation Group. It is based on the
procedure of dynamical hadronization, which comes with the great
advantage that the information of bound states can be mapped to lower
order $n$-point correlation functions in a systematic
manner. Furthermore, this simplifies the procedure of building
self-consistent truncations, as all correlation functions are
generated from a single master equation, and external input can be
incorporated without further problems.  These advantages circumvent
many of the problems faced when expressing the bound states
exclusively in terms of their constituents.

This framework was applied to the pion and sigma mesons within a
quark-meson model motivated via dynamical hadronization from QCD and
solving the corresponding flow equation for Euclidean momenta. In
order to test the validity of our truncation we have compared the
common subset of results with a recent study of the Euclidean system
in first principle QCD \cite{Cyrol:2017ewj}. While we found at low
energies good quantitative agreement, above scales of
$\gtrsim \unit[250]{MeV}$ qualitative deviations start to
appear. These deviations could be traced back to missing contributions
in the Yukawa coupling.

The bound state properties of the pion and the sigma meson were
accessed from the Euclidean correlators via a suitable Pad{\'e}
approximation extracting hereby the pole masses. While the pole mass
of the pion, the lowest lying excitation, agrees very well with the
expectation from the Euclidean curvature mass, we found not
unexpectedly a significant deviation of $\sim 35\%$ for the sigma
meson.

Most importantly, the here presented approach can and will be
systematically extended towards the full bound state spectrum of QCD.


\section{Acknowledgments}
\label{Acknowledgments}

J.\ P.-L. and W.\ A.\ M.\ are supported by the FWF Doctoral Program
W1203 "Hadrons in vacuum, nuclei and stars" and by the mobility
program of the Science Faculty of the University of Graz.  M.\ M.\ is
supported by the FWF grant J3507-N27, the DFG grant MI 2240/1-1 and
the U.S. Department of Energy under contract de-sc0012704. This work
is also supported by EMMI, the BMBF grant 05P18VHFCA, and is part of
and supported by the DFG Collaborative Research Centre SFB 1225
(ISO-QUANT).


\bookmarksetup{startatroot}
\appendix

\section{Additional definitions and technical details}
\label{app:add_details}
In this section additional informations regarding our ansatz for the
effective action \labelcref{eq:qm_model} and its solution are
provided.  Furthermore some additional definitions are introduced,
used to keep the flow equations as simple as possible.

\subsection{Regulator and propagators}
The momentum independent part of the mesonic two-point function can be
obtained from the effective potential
\begin{align}\nonumber 
\Gamma^{(2)}_{\vec{\pi}}(p=0) &= \partial_{\rho} V(\rho)\,, \\[1ex]
\Gamma^{(2)}_\sigma(p=0) &= \partial_{\rho} V(\rho) + 2 \rho\, \partial_{\rho}^2 V(\rho)
\, ,
\end{align}
which can be obtained from matching the corresponding flow equations.
Instead of solving the flow for the full mesonic two-point functions
only the momentum dependent part needs to be considered:
\begin{align}
\label{eq:delta_prop}
  \Delta\Gamma^{(2)}_{i,k}(p) = \Gamma^{(2)}_{i,k}(p) - \Gamma^{(2)}_{i,k}(0) \, ,
  \quad i\in\{\sigma,\vec{\pi}\}
  \, .
\end{align}
For the flows of the momentum dependent part only the polarization
diagrams have to be calculated, as the four-point functions do not
carry any momentum dependence in our truncation and all tadpoles
vanish as a result.  The mesonic wave-function renormalizations are
directly related to the two-point functions by
\begin{align}
Z_{i,k}(p) = p^{-2} \Delta\Gamma^{(2)}_{i,k}(p) \, .
\end{align}
The fermion mass and the Yukawa coupling are related by
\begin{align}
m_{q,k}^2(p) = \frac{\rho}{N_f} h_k(p,-p)
\, .
\end{align}
As a regulator shape function we are using a standard exponential regulator
\begin{align}\nonumber 
r_B(x) &= \frac{x^{m-1}}{e^{x^m}-1}\,, \\[1ex]
r_F(x) &+ 1 = \sqrt{r_B(x) + 1} 
\, ,
\end{align}
resulting in the following full regulators
\begin{align}\nonumber 
  R_{i,k}(p) &= Z_{i,k}(0) p^2\, r_B\!\left(\frac{p^2}{k^2}\right) \, ,
               \quad i\in\{\sigma,\vec{\pi}\} ~,,\\[1ex]
  R_F(p) &= \hat{R}(p) \slashed{p} \, \quad \hat{R}(p) =
           Z_{q,k}(0) r_F\!\left(\frac{p^2}{k^2}\right)
           \, .
\end{align}
We define the following mesonic propagators
$(i\in\{\sigma,\vec{\pi}\})$
\begin{align}
 G_{i}(p)=& \frac{1}{ Z_{i,k}(p)p^2 + R_{i,k}(p) +
 \Gamma^{(2)}_{i}(0)} \, .
\end{align}
In addition we can define an effective (scalar) quark propagator,
entering all loop functions
\begin{align}\nonumber 
  G_q (p)&= \frac{1}{\left[Z_{rq,k}(p)\right]^2 + m_q^2(p)} \, ,
  \\[1ex]
  Z_{rq,k}(p) &= Z_{q,k}(p)+ \hat{R}(p)
                \, .
\end{align}

\subsection{Effective potential}
The effective potential $V_k(\rho)$ is solved using a Taylor expansion
about the IR minimum of the potential including the explicit symmetry
breaking term. The bare expansion point is kept fix, therefore greatly
increasing the stability of the equations \cite{Pawlowski:2014zaa}. It
has been verified that the results are independent of the order of the
Taylor expansion.

\subsection{Initial values}
Our initial values are chosen such that the physical values for the
quark mass $\bar{m}_q=\unit[298]{MeV}$, the expectation value of the
chiral condensate $\sigma_{\text{\tiny{min}}}=\unit[93]{MeV}$, the
pion mass $\bar{m}_{\vec{\pi}}=\unit[139]{MeV}$ and the curvature mass
of the sigma meson $\bar{m}_\sigma=\unit[495]{MeV}$ are obtained.
Since the sigma meson is a scattering state in this model, it is not
possible to directly assign it a pole mass, therefore we resorted to
the curvature mass, which we fixed slightly higher than the physical
mass. More details on the relation between curvature and pole masses,
can be found in \Cref{sec:result_minkowski} or e.g.\
\cite{Pawlowski:2017gxj}.  The flow is initiated at
$\Lambda_\text{\tiny{UV}}=\unit[950]{MeV}$, in accordance with our
discussion in \Cref{subsec:QCDtoQM}.  The initial potential is assumed
to be quadratic
\begin{align}
  V_{k=\Lambda_\text{\tiny{UV}}} = a_1 (\rho-\rho_0) + \
  \frac{a_2}{2}(\rho-\rho_0)^2
  \, ,
\end{align}
where $\rho_0$ is the expansion point. All other orders of the
effective potential are zero.  The wave-function renormalizations are
set to unity at the UV cutoff. For completeness we also state our
initial values for our full truncation considered in this work
\begin{align}\nonumber 
a_1 &= (2638 \text{ MeV})^{2} \\[1ex]\nonumber 
a_2 &= 50 \\[1ex]\nonumber 
h_{\Lambda} & = 18.085 \\[1ex]
\rho_{0} &= 28.32 \text{ MeV}
\, .
\end{align}

\subsection{Dynamical Hadronization}
 \label{app:DH}
 Our dynamical hadronization procedure follows \cite{Cyrol:2017ewj}.
 To this end we parametrize the scale dependence of the auxiliary field as
 \begin{align}\nonumber 
 \dot{\sigma}(p) =& \frac{1}{\sqrt{2 N_f}} \int_{q} \dot{A}\left(p-q,q\right) 
 \overline{q}(p-q) q(q)
 \\[1ex]
 \dot{\vec{\pi}}(p) =& \int_{q} \dot{A}\left(p-q,q\right) 
 \overline{q}(p-q) \imag \gamma_5 \vec{\tau} q(q)
 \, .
\end{align}
In \cite{Cyrol:2017ewj} it was found that a weighted sum between total
and relative momenta captures the full momentum dependence quite
accurately
\begin{align}
 \dot{A}(p_1,p_2) &= \dot{A}\left( (p_1+p_2)^2+\frac{1}{4}(p_1-p_2)^2\right)
 \, .
\end{align}
As a result the flow of the corresponding channel in the four-Fermi
interaction $\lambda_k$ is set to zero
\begin{align}\nonumber 
 \partial_k \lambda_k(p,-p,p) =& \text{Flow} \lambda_k(p,-p,p)
 - \dot{A}(p^2) h_k (p^2) \\[1ex]
 \dot{A}(p^2) =& \frac{\text{Flow} \lambda_k(p,-p,p)}{h_k (p^2)}
 \, .
\end{align}

\begin{figure*}[t!]
	\includegraphics{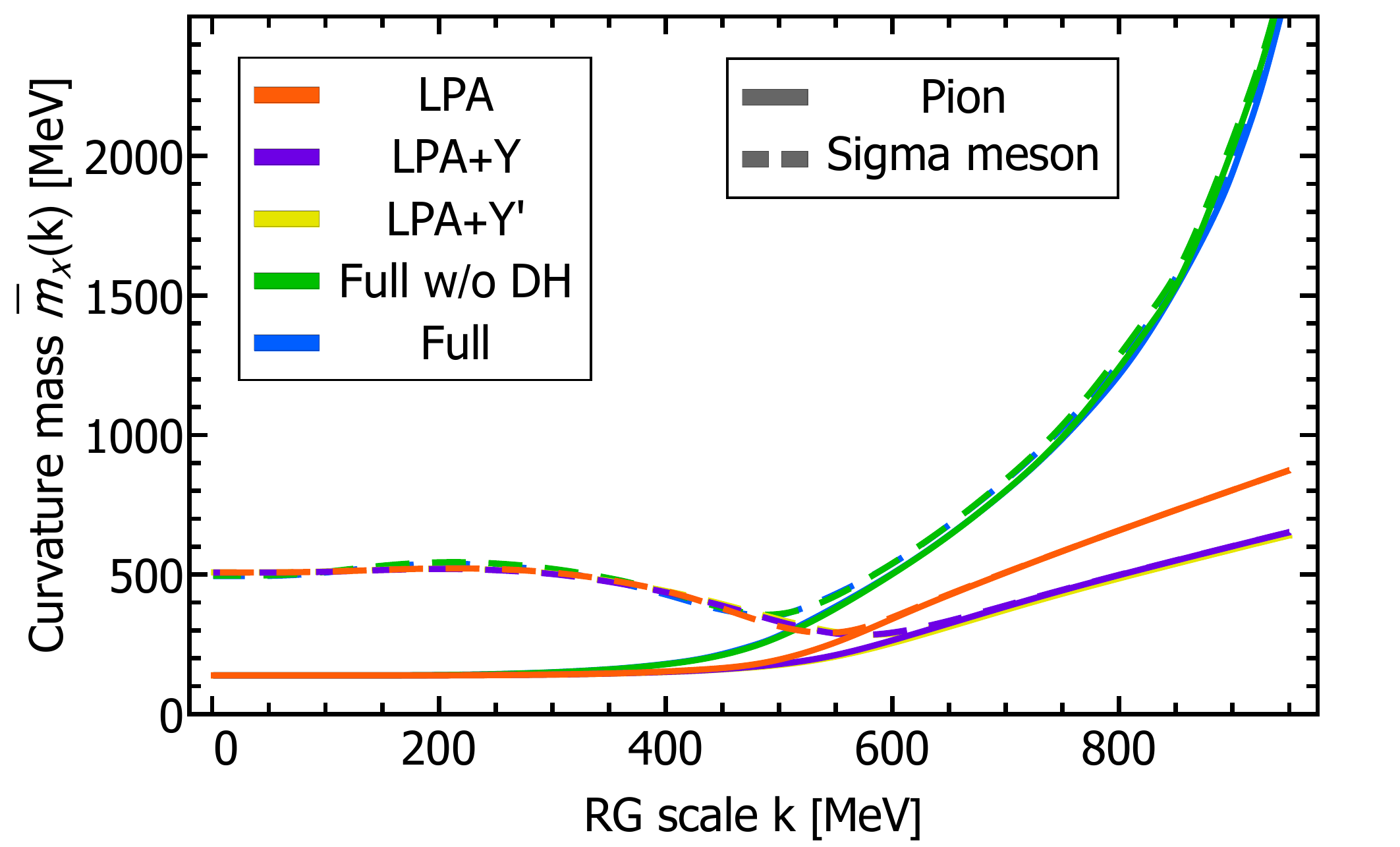}\hfill
	\includegraphics{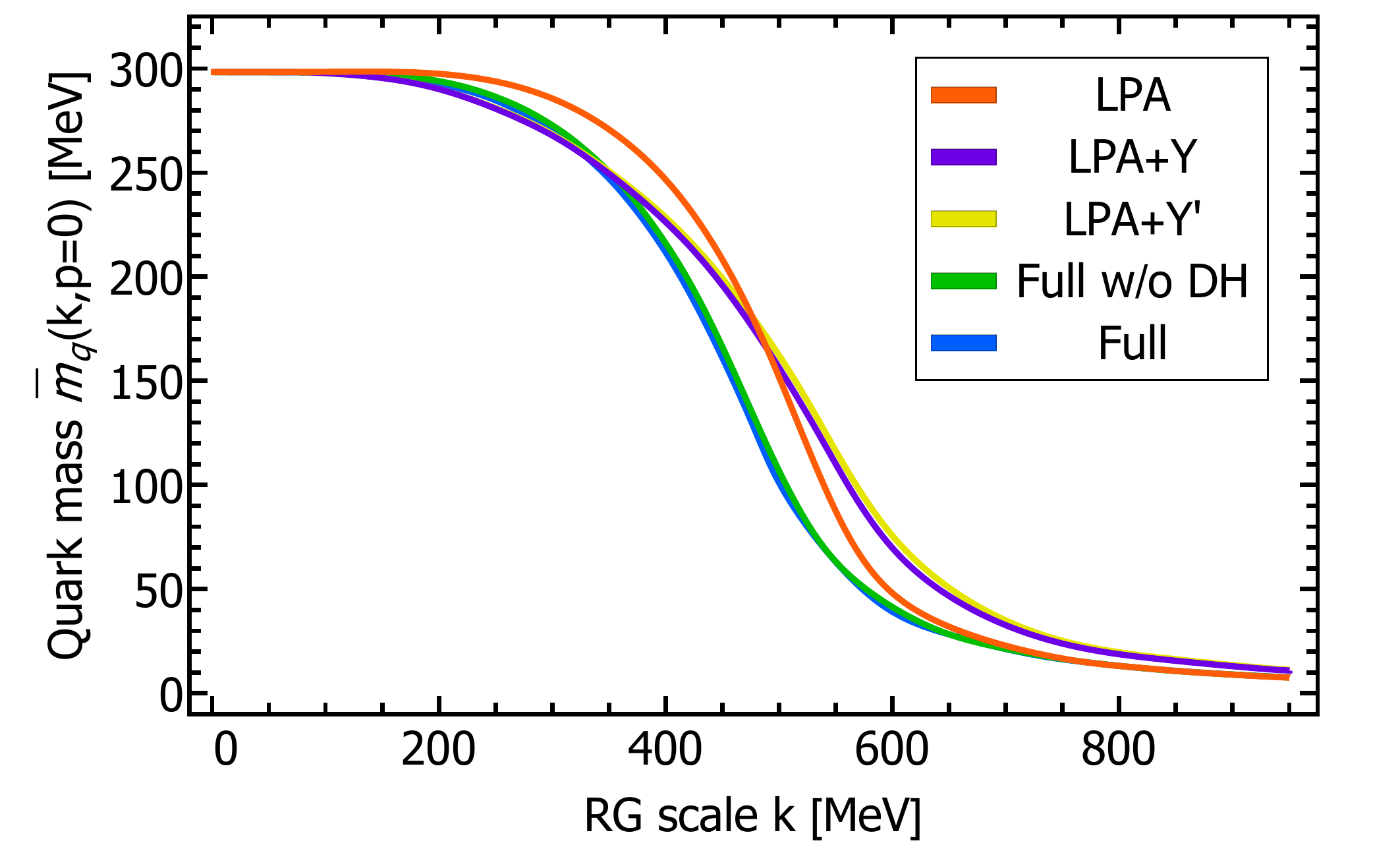}
	\caption{Flow of the meson masses (left panel) and quark
          masses at $p=0$ (right panel), shown for different
          truncations.}
	\label{fig:truncations_masses}
\end{figure*}

\begin{figure}[t!]
	\includegraphics[width=0.5\textwidth]{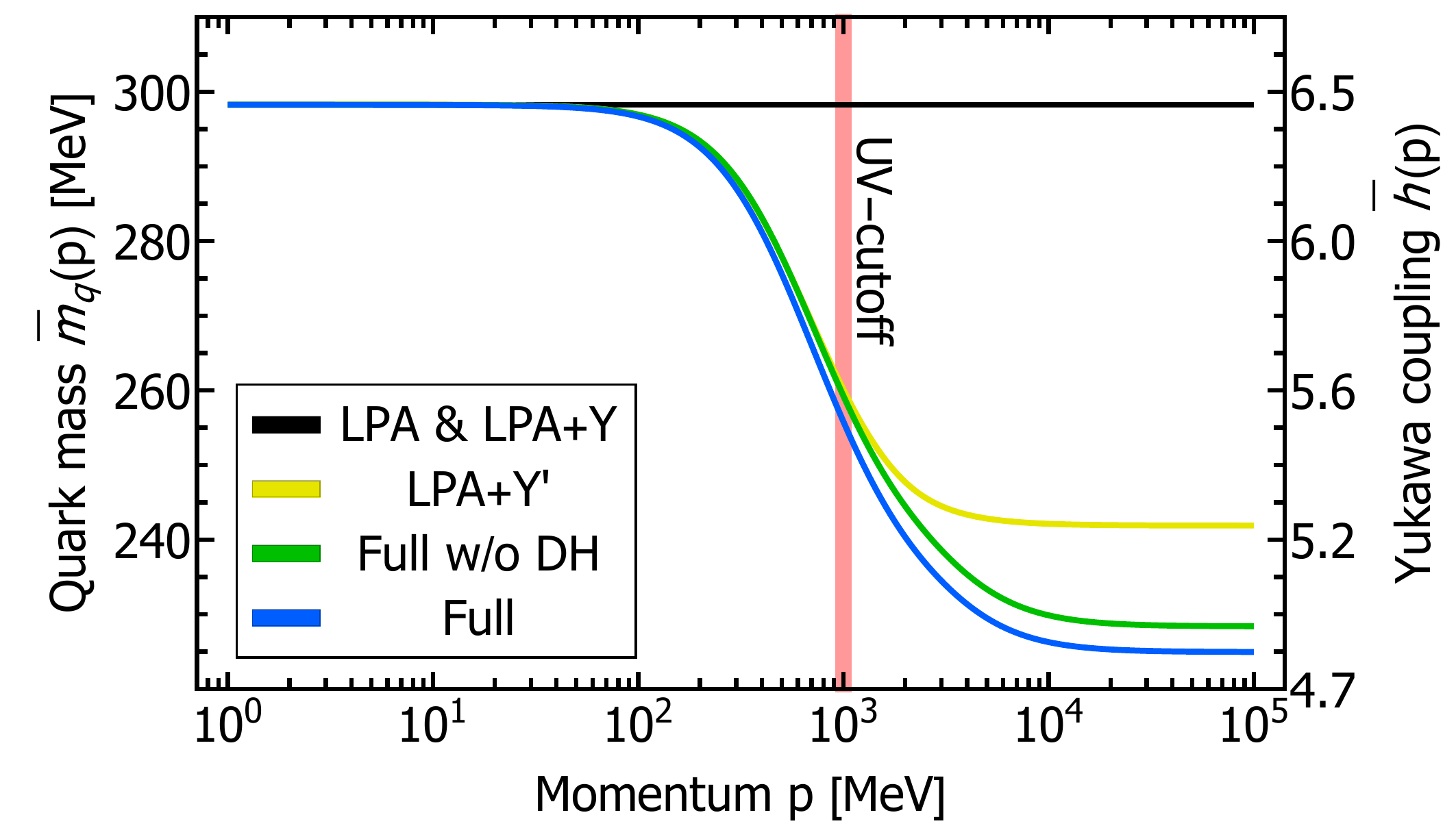}
	\caption{The momentum dependence of the quark mass (left axes)
          and Yukawa coupling (right axes) for $k=0$ and different
          truncations is depicted. The vertical red line shows the
          value of the UV cutoff.}
	\label{fig:truncations_yukawa}
\end{figure}

\begin{figure*}[t!]
	\includegraphics{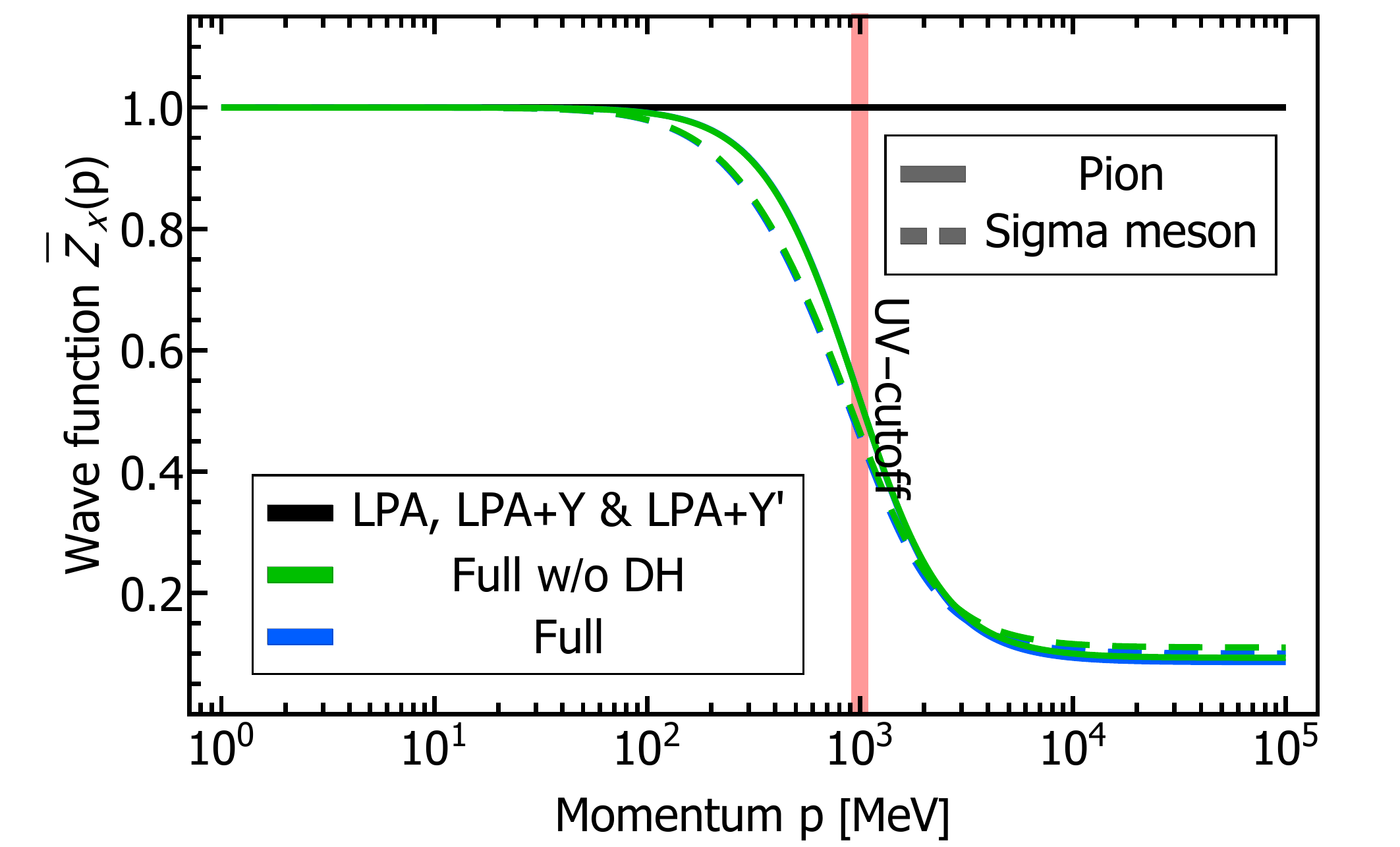}\hfill
	\includegraphics{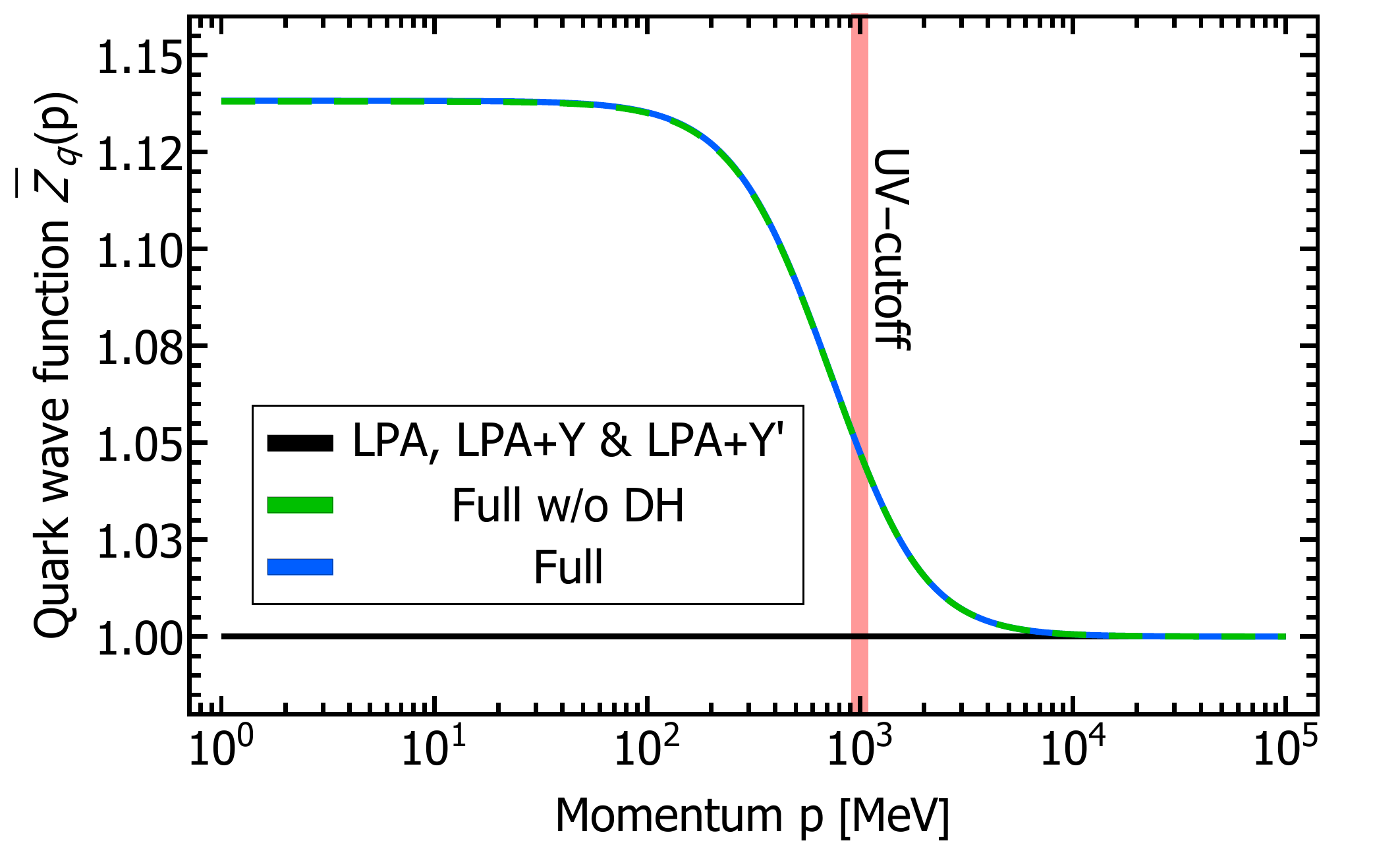}
	\caption{The momentum-dependent wave function renormalizations
          for the meson (left panel) and quarks (right panel) is
          plotted for $k=0$ and different truncations.}
	\label{fig:truncations_wavefunctions}
\end{figure*}

\section{Flow equations}
 \label{app:flow_equations}
 For the derivation of the flow equations DoFun \cite{Huber:2011qr}
 was used. For the subsequent tracing of the equations we have used
 FormTracer \cite{Cyrol:2016zqb}.  We do not state the tadpole (TP)
 contributions explicitly, since we obtain the momentum independent
 part from \labelcref{eq:delta_prop}.  Let us define the relative
 momenta between the internal and the external momenta $r=q-p$.  The
 contribution to the $\dot{\Gamma}^{(2)}_{\pi,k}(p)$ and
 $\dot{\Gamma}^{(2)}_{\sigma,k}(p)$ from the quark loop is the same,
 which is given by
\begin{align}
 \dot{\Gamma}^{(2),q}_{\phi,k}(p) = & \int_{q} \left\{
 12 \dot{\hat{R}}_k^{q}(q) G^2_{q}(q) G_{q}(r) h_{k}(r,-q) h_{k}(q,-r) 
\right. \nonumber \\[1ex] & 
   \left[ \left(Z^2_{rq,k}(q) q^2 - m^2_{q,k}(q)  \right)
   Z_{rq,k}(r) (p\,q - p^2)  
\right. \nonumber \\[1ex]  & \left. \left.
  +  2 m_{q,k}(q) m_{q,k}(r) Z_{rq,k}(q) q^2 \right] \right\} \, .
\end{align}
The full flow for the two-point function of the pion is 
\begin{align}
 \dot{\Gamma}^{(2)}_{\pi,k}(p) = & \dot{\Gamma}^{(2),q}_{\phi,k}(p) + \int_{q} \left\{
  \left(\Gamma^{(3)}_{\pi\pi\sigma}(0) \right)^2  \left[ \dot{R}_{\pi,k}(q) G^2_{\pi}(q)    
  \right. \right. \nonumber \\[1ex] & \left. \left.
  G_{\sigma}(r) + \dot{R}_{\sigma,k}(q) G^2_{\sigma}(q) G_{\pi}(r) \right]
  \right\} + \text{TP}
  \,  ,
\end{align}
and correspondingly the full flow of the sigma is
\begin{align}
 \dot{\Gamma}^{(2)}_{\sigma,k}(p) = & \dot{\Gamma}^{(2),q}_{\phi,k}(p) + \int_{q} \left\{
  3 \left( \Gamma^{(3)}_{\pi\pi\sigma}(0) \right)^2 \dot{R}_{\pi,k}(q) G^2_{\pi}(q) G_{\pi}(r)   
  \right. \nonumber \\[1ex] & \left.
  + \left( \Gamma^{(3)}_{\sigma \sigma \sigma}(0) \right)^2 \dot{R}_{\sigma,k}(q)
   G^2_{\sigma}(q) G_{\sigma}(r)
  \right\} + \text{TP}
  \, .
\end{align}
The mesonic three point vertices are obtained from the effective potential 
\begin{align}\nonumber 
  \Gamma^{(3)}_{\pi\pi\sigma}(0) &= \left. \sigma V_k^{(2)}[\rho]
                                   \right|_{\rho=\rho_0} \, ,\\[1ex]
  \Gamma^{(3)}_{\sigma\sigma\sigma}(0) &= \left. \left[  3 \sigma V_k^{(2)}[\rho]
                                         + \sigma^{3} V_k^{(3)}[\rho] \right] \right|_{\rho=\rho_0} \, .
\end{align}
The flow for the wave function part of the quark propagator reads
\begin{align}\nonumber 
 \dot{Z}_{q,k}(p) =& -\frac{1}{4 p^2} \int_{q} \left\{
 (p\,q) \dot{\hat{R}}_k^{q}(q) G^2_{q}(q) h_{k}(-p,q) h_{k}(-q,p)
 \right. \nonumber \\[1ex] & 
 \left(m^2_{q,k}(q) - Z^2_{rq,k}(q) q^2 \right) 
 \left( 3 G_{\pi}(r) + G_{\sigma}(r)\right)
 \nonumber \\[1ex] &
 +2 G_{q}(r) Z_{rq,k}(r) (p\, q - p^2)  h_{k}(r,p) h_{k}(-p,-r)
 \nonumber \\[1ex] & \left.
                \left( 3 \dot{R}_{\pi,k}(q) G^2_{\pi}(q)+  \dot{R}_{\sigma,k}(q)
                G^2_{\sigma}(q)  \right)  \right\} + \text{TP}
 \, ,
\end{align}
and finally the flow of the Yukawa coupling:
\begin{align}
   \dot{h}_{k}(p,-p) =& -\frac{1}{4} \int_{q} \left\{
   2 \dot{\hat{R}}_k^{q}(q) G^2_{q}(q) h_{k}(-p,-q) h_{k}(q,p)
   \right. \nonumber \\[1ex] &
   h_{k}(q,-q) Z_{rq,k}(q) q^2 \left( 3 G_{\pi}(r) - G_{\sigma}(r)\right)
   \nonumber \\[2ex] & 
   + G_{q}(r) h_{k}(-p,-r) h_{k}(r,p)  h_{k}(r,-r)
    \nonumber \\[1ex] & \left.
    \left( 3 \dot{R}_{\pi,k}(q) G^2_{\pi}(q)- \dot{R}_{\sigma,k}(q)  G^2_{\sigma}(q)
    \right) \right\}
    \nonumber \\[1ex] &
    -\dot{A}(p,-p) \frac{\Gamma^{(1)}_{\phi}(0)}{\sigma}
    \, .
\label{Eqn:Flow_Yuk}\end{align}
The flow of $\lambda_{k}$ has a very lengthy expression and thus is not noted here.

\section{Truncations}
 \label{Truncations}
 
 In this section of the Appendix we discuss details on the different
 truncations analysed within the quark-meson model and compare them to
 the full case. All truncations are set to get the same IR
 physics. The results are shown in \Cref{fig:truncations_masses},
 \Cref{fig:truncations_yukawa} and
 \Cref{fig:truncations_wavefunctions}.
 
 In total we consider four additional truncations
 \begin{itemize}
 \item LPA: The wave functions are set to unity, the Yukawa coupling
   is fixed to a constant and the rebosonisation is ignored.
 \item LPA + Y: The $k$-dependence of the Yukawa coupling is
   additionally taken into account.
 	\item LPA + Y': The momentum dependence of the Yukawa coupling
          is additionally taken into account.
 	\item Full w/o DH: The full wave functions are taken
          additionally into account, this corresponds to the full
          truncation without the rebosonisation procedure.
 \end{itemize}

 \noindent Turning to the running of the curvature masses, shown in
 \Cref{fig:truncations_masses}, the most notable effect is the faster
 decoupling of the mesons at larger scales, hence getting closer to
 the behaviour in full QCD, while the running of the quark mass is
 mostly unaffected. The momentum dependence of the Yukawa coupling,
 shown in \Cref{fig:truncations_yukawa}, is below the UV-cutoff
 essentially independent of the truncation as long as it is
 calculated. A similar result is found for the wave functions, shown
 in \Cref{fig:truncations_wavefunctions}. These findings fit to the
 discussion about the different contributions generating momentum
 dependencies in \Cref{sec:toQCD}.

\bibliography{references}

\end{document}